\documentclass[onecolumn]{IEEEtran}
\usepackage[utf8]{inputenc}
\usepackage{amsmath,bm}
\usepackage{amsthm}
\usepackage{amssymb}
\allowdisplaybreaks
\usepackage{graphicx}
\usepackage{xcolor}
\usepackage[english]{babel}
\usepackage{tabularx}
\usepackage{multirow}
\usepackage{tikz}\usetikzlibrary{shapes.multipart}
\usetikzlibrary{positioning}
\tikzset{font={\fontsize{9pt}{12}\selectfont}}
\tikzset{>=latex}
\usetikzlibrary{calc}
\usetikzlibrary{arrows}
\usetikzlibrary{external}
\usetikzlibrary{patterns}
\usetikzlibrary{fit}
\usepackage[final]{changes}

\newcommand{\rcomment}[1]{\textcolor{red}{#1}}

\newcommand{\eqdef}{\triangleq}

\DeclareMathOperator{\rt}{drt}
\DeclareMathOperator{\wt}{wt}
\DeclareMathOperator{\ir}{Irr}
\DeclareMathOperator{\rl}{RLL}
\DeclareMathOperator{\ins}{Sp}
\DeclareMathOperator{\inl}{IL}
\DeclareMathOperator{\cusum}{CS}
\DeclareMathOperator{\Od}{Od}
\DeclareMathOperator{\Even}{Ev}
\DeclareMathOperator{\OdEv}{OE}

\newtheorem{example}{Example$\!$}
\newtheorem{theorem}{Theorem$\!$}

\newtheorem{construction}{Construction$\!$}
\usepackage{graphicx}

\usepackage{xcolor}

\newcommand{\bcomment}[1]{\textcolor{blue}{#1}}

\newcommand{\cC}{\mathcal{C}}

\newcommand{\mybold}[1]{\bm{#1}}

\newcommand{\bc}{{\mybold{c}}}

\newcommand{\be}{{\mybold{e}}}

\newcommand{\br}{{\mybold{r}}}
\newcommand{\bs}{{\mybold{s}}}

\newcommand{\bu}{{\mybold{u}}}
\newcommand{\bv}{{\mybold{v}}}
\newcommand{\bw}{{\mybold{w}}}
\newcommand{\bx}{{\mybold{x}}}
\newcommand{\by}{{\mybold{y}}}
\newcommand{\bz}{{\mybold{z}}}

\newcommand{\bepsilon}	{\mybold{\epsilon}}

\newcommand{\bmu}		{\mybold{\mu}}

\begin{document}
\title{Error-correcting Codes\\ for Noisy Duplication Channels}
\author{
  \IEEEauthorblockN{Yuanyuan Tang 
                     and Farzad Farnoud
                     }
                     \\
  \IEEEauthorblockA{
                    Electrical \& Computer Engineering,
                    University of Virginia,
                    \texttt{\{yt5tz,farzad\}@virginia.edu}} \\
  \thanks{This work was supported in part by NSF grants under grant nos.~1816409 and~1755773
  . This paper was presented in part at the 57th Annual Allerton Conference on Communication, Control, and Computing in 2019 \cite{Yuanyuan2019Error}.}
}

\maketitle

\begin{abstract}
Because of its high data density and longevity, DNA is emerging as a promising candidate for satisfying increasing data storage needs. Compared to conventional storage media, however, data stored in DNA is subject to a wider range of errors resulting from various processes involved in the data storage pipeline. In this paper, we consider correcting duplication errors for both exact and noisy tandem duplications of a given length $k$. An exact duplication inserts a copy of a substring of length $k$ of the sequence immediately after that substring, e.g., $\mathsf{ACGT} \to \mathsf{ACG\underline{ACG}T}$, where $k=3$, while a noisy duplication inserts a copy suffering from substitution noise, e.g., $\mathsf{ACGT} \to \mathsf{ACG\underline{A\rcomment{T}G}T}$. Specifically, we design codes that can correct any number of exact duplication and one noisy duplication errors, where in the noisy duplication case the copy is at Hamming distance 1 from the original. Our constructions rely upon recovering the duplication root of the stored codeword. We characterize the ways in which duplication errors manifest in the root of affected sequences and design efficient codes for correcting these error patterns. We show that the proposed construction is asymptotically optimal, in the sense that it has the same asymptotic rate as optimal codes correcting exact duplications only.
\end{abstract}


\section{Introduction}
The rapidly increasing amount of data and the need for long-term data storage have led to new challenges. 
In recent years, advances in DNA sequencing, synthesis, and editing technologies~\cite{yazdi2015dna,tang2019single} have made deoxyribonucleic acid (DNA) a promising alternative to conventional storage media. Compared to traditional media, DNA has several advantages, including high data density, longevity, and ease of copying information. For example, it may be possible to recover a DNA sequence after $10,000$ years and a single human cell contains an amount of DNA that can ideally hold $6.4$ Gb of information~\cite{yazdi2015dna}. Research in the past few years has led to significant advances, such as the ability to provide random-access to the data~\cite{yazdi2015rewritable} as well as a DNA storage system with portable size~\cite{yazdi2017portable}. 
Nevertheless, there are still significant challenges to be overcome. One obvious challenge is that a diverse set of errors is possible, including substitutions, duplications, insertions, and deletions. 
This paper focuses on error-correcting codes for noisy duplication channels. In such case, in addition to exact duplications, a noisy duplication, where an approximate copy is inserted into the sequence, may occur.

In duplication channels, (tandem) duplication errors generate copies of substrings of the sequence and insert each copy after the original substring~\cite{jain2017duplication}. This type of channel was first studied in the context of recovering from timing errors in communication systems that led to individual symbols being repeated~\cite{dolecek2010repetition}. The copying mechanism of DNA, however, allows multiple symbols being repeated, for example, via slipped-strand mispairings, where the slippage of the molecule copying DNA causes a substring to be repeated~\cite{jain2017duplication}. Properties of duplication in DNA have been studied from various vantage points, including the theory of formal languages and the entropy of DNA sequences (see, e.g.,~\cite{lou2018} and references therein). 
Codes for correcting duplication errors in the context of data storage in the DNA of living organisms, such as bacteria~\cite{shipman2017}, were studied by~\cite{jain2017duplication}, where optimal constructions for correcting exact duplications of constant length were presented. This and related problems were then further studied by a number of works including~\cite{jain2017noise,yehezkeally2018reconstruction,kovavcevic2018asymptotically,lenz2019duplication,chee2017,tang2019single}. Most related to this paper is~\cite{tang2019single}, which studies error correction in duplication and substitution channels, when substitutions are independent from duplications and when they only occur in copies generated by duplications. The latter model, i.e., the \textit{noisy duplication model}, which is motivated by the abundance of inexact copies in tandem repeat stretches in genomes~\cite{pumpernik2008replication}, is the model studied in this work. 

In the noisy duplication channel, two types of errors are possible: i) exact duplications, which insert an exact copy of a substring in tandem, such as $\mathsf{ACGTC}\to \mathsf{ACGT\underline{CGT}C}$; and ii) noisy duplications, which insert approximate copies, e.g., $\mathsf{ACGTC}\to \mathsf{ACGT\underline{C\rcomment{T}T}C}$. In both cases, the length of the duplication refers to the length of the duplicated substring (3 in our preceding examples). In this paper, we limit our attention to exact and noisy tandem duplications of length $k$, referred to as $k$-TDs and $k$-NDs, respectively.  Furthermore, we only consider noisy duplications where the copy and the original substring differ in one position. In other words, each noisy duplication can be viewed as an exact duplication followed by a substitution in the inserted copy. One may also consider \textit{left} duplications, which add the copy before the substring, as opposed to \textit{right} duplications discussed above. Left and right duplications are the same for exact duplications but not for noisy duplications. If the direction is known, codes proposed here for right duplications can be applied to left duplications by reversing the sequence. Stronger codes are needed to correct the errors if the direction is unknown.

We will design codes that correct (infinitely) many $k$-TD and a single $k$-ND errors, as a step towards codes that can correct $t_1$ $k$-TDs and $t_2$ $k$-NDs, for given $t_1$ and $t_2$. 
The proposed codes will rely on finding the duplication root of the stored codeword. The \textit{duplication root} of a sequence $\bx$ is the sequence obtained from $\bx$ by removing all repeats of length $k$. While $k$-TDs do not alter the duplication root, $k$-NDs do. Thus, we will first analyze the effect of noisy duplications on the root of the sequence. We show that the root may change in a variety of ways, leading to several error patterns. We then design efficient error-correcting codes that correct these errors via a number of transforms that simplify the different error patterns.

For codes capable of correcting any number of exact duplications, the best possible asymptotic rate (i.e., the limit of the rate for code length $n\to\infty$, as defined in~\eqref{eq:cap-def}) was given in~\cite{jain2017duplication} as
\begin{equation}\label{eq:kTDrate}
    1-\frac{(q-1)\log_q e}{q^{k+2}} + o(q^{-k}),
\end{equation}
where $o(q^{-k})$ represents terms whose ratio to $q^{-k}$ vanishes as $k$ becomes larger. The question arises as to whether it is possible to correct an additional noisy duplication without a rate penalty. It is worth noting that the best known code for correcting an additional unrestricted substitution, i.e., a substitution that can occur anywhere rather than in a copy generated by duplication, has rate that is bounded from below by~\cite{tang2019single}
\begin{equation}
    1 - \frac{2}{k} \log_q\frac{q}{q-1} + o(1).
\end{equation}
which indicates a rate penalty. In contrast, we show that the proposed codes have the same asymptotic rate as~\eqref{eq:kTDrate}, and are thus asymptotically optimal.

We note that for $q=4$, which is of interest in DNA storage, and for $k=2,3,4$, the asymptotic rates of optimal codes correcting any number of exact $k$-TDs can be shown to equal 0.9613, 0.9912, 0.9979, respectively~\cite{jain2017duplication}. The fact that these values are close to 1 indicates that the rate penalty for correcting an infinite number of exact duplication errors compared to only correcting a finite number is not significant and diminishes as $k$ grows. In this work, we have focused on correcting many exact duplications rather than a finite number.
\deleted{
In this paper, we consider noisy duplication channels that contain at most one substitution error in one of many tandem duplication copies. In the noisy duplication channel, we consider all the tandem duplication errors of length $k$ ($K$-TDs). In addition to $k$-TDs, a substitution of one symbol may occur in one copy of $k$-TDs. Since all the exact $k$-TDs keep irreducible strings of sequences \cite{jain2017duplication}, We consider codes that can correct all the errors over irreducible strings caused by many exact \bcomment{$k$-TDs} and at most one substitution error. After analyzing all the changes over the irreducible strings, the paper proposes an error correcting code for noisy duplication channels with a lower bound of the code size. }

This paper is organized as follows. The notation and preliminaries are given in Section~\ref{sec:Not_Pre}. In Section~\ref{sec:descands_noi_channel}, we analyze the error patterns that manifest as the result of passing through the noisy duplication channel. Finally, the code construction and the corresponding code size are presented in Section~\ref{sec:correct_code}. \deleted{ Note this}

\section{Notation and Preliminaries}\label{sec:Not_Pre}

Throughout the paper, $\Sigma_q$ represents a finite alphabet of size $q$, \added{assumed without loss of generality to be $\{0,1,\dotsc,q-1\}$. We use $\Sigma^+_q$ to denote the nonzero elements of $\Sigma_q$ and $\Sigma_q^{*}$ to denote all strings of finite length over $\Sigma_q$. In particular, $\Sigma_q^*$ includes the empty string $\Lambda$.} Furthermore, $\Sigma_q^{n}$ represents the strings of length $n$ over $\Sigma_q$ and $\Sigma_q^{\le n}$ is the set of strings of length at most $n$. \added{The set $\{1,\dotsc,n\}$ is represented by $[n]$.}

We use bold symbols, such as $\bx$ and $\by_j$, to denote strings over $\Sigma_q$. The entries of strings are shown with normal symbols, e.g., $\bx=x_1x_2\dotsm x_n$ and $\by_j=y_{j1}y_{j2}\dotsm y_{jm}$\added{, where $x_i,y_{ji}\in\Sigma_q$}. \replaced{The indices of elements }{Symbol indexes }of words over $\Sigma^{*}_{q}$ start from $1$, unless otherwise \replaced{stated}{started}. \deleted{For example, a word $\bx\in \Sigma^{n}_{q}$ can be denoted as $\bx=x_1x_2\cdots x_n$, where $x_{i}\in \Sigma_{q}$.}For two words $\bx, \by\in \Sigma_q^{*}$, \replaced{their concatenation} {the concatenation of them} is denoted as $\bx\by$, and $\bx^{m}$ represents \replaced{the }{a} concatenation of $m$ copies of $\bx$. Given a word $\bx\in \Sigma_q^{*}$, the length of $\bx$ is represented as $|\bx|$. In addition, for a word $\bx\in \Sigma^{*}_{q}$, the Hamming weight $\wt(\bx)$ denotes the number of non-zero symbols in $\bx$. If a word $\bx\in \Sigma_{q}^{*}$ can be expressed as $\bx=\bu\bv\bw$ with $\bu,\bv,\bw \in \Sigma_{q}^{*}$, then $\bv$ is a substring of $\bx$.

Given a word $\bx\in \Sigma_{q}^{*}$, an \added{(exact)} \emph{tandem duplication} of length $k$ ($k$-TD) generates a copy of a substring \added{$\bv$} of $\bx$ of length $k$ and inserts the copy \replaced{immediately after $\bv$}{directly following that substring}. More specifically, a $k$-TD can be expressed as \cite{jain2017duplication}
\begin{equation}\label{eq:k_TD}
    T_{i,k}(\bx)=
    \begin{cases}
           \bu\bv\bv\bw & \text{if } \bx=\bu\bv\bw, |\bu|=i, |\bv|=k   \\
           \bx & \text{if } |\bx|<i+k
    \end{cases}
\end{equation}
For example, given \replaced{the}{an} alphabet $\Sigma_{3}=\{0,1,2\}$ and $k=3$, \replaced{a}{one} $k$-TD \replaced{may result in}{can be expressed as}
\begin{equation}\label{eq:ex0_k_TD}
    \begin{split}
        \bx=1201210 \rightarrow \bx'=T_{1,3}(\bx)=1201\underline{201}210,
    \end{split}
\end{equation}
where the underline\added{d} substring $201$ is the copy. \added{We refer to} $\bx'$ \replaced{as a $k$-TD \emph{descendant}}{is a $k$-TD descendant} of $\bx$.

Given a word $\bx\in \Sigma_{q}^{n}$ \added{$n\ge k$}, the $k$-\emph{discrete-derivative} transform~\cite{jain2017duplication} is defined as $\phi(\bx)=(\hat\phi(\bx),\bar\phi(\bx))$, where
 \begin{equation}\label{eq:k_decre_deriv}
    \hat\phi(\bx)=x_1\cdots x_k, \bar\phi(\bx)=x_{k+1}\cdots x_{n}-x_1 \cdots x_{n-k}.
\end{equation}
where \deleted{$n \geq k$, and}the subtraction is performed entry-wise \replaced{modulo $q$}{over $\Sigma_{q}$}. \replaced{Continuing the example given in~\eqref{eq:ex0_k_TD} with $k=3$ and $q=3$,}{Based on the transform~\eqref{eq:k_decre_deriv}, the $k$-TD example~\eqref{eq:ex0_k_TD} can be expressed as }
 \begin{equation}\label{eq:ex1_k_TD_decre}
    \begin{split}
         \bx=1201210 &\rightarrow \bx'=1201\underline{201}210,\\
        \phi(\bx) = 120,0012 &\rightarrow \phi(\bx') = 120,0\underline{000}012.
    \end{split}
\end{equation}
As seen in the example\deleted{ above}, after \replaced{the}{one} $k$-TD\deleted{ error} in $\bx$, \deleted{the} $\bar\phi(\bx')$ can be obtained by inserting $0^{k}$ into $\bar\phi(\bx)$, immediately after the $i$-th entry.

 \replaced{Copies generated by tandem duplications}{duplication copies of the $k$-TDs} may not be always perfect. \added{That is, the copy may not always be exact. Such a duplication is referred to as a \emph{noisy duplication}.} \replaced{In this paper, we limit our attention to noisy duplications in which the copy is at Hamming distance 1 from the original.}{If the noise is only added to one of the $k$ symbols of one copy, we call it a \emph{noisy-duplication}.} Continuing example~\eqref{eq:ex0_k_TD}, one symbol in the copy $201$ may change,
 \begin{equation*}
    \begin{split}\label{eq:ex1_k_TD_sub_decre}
         \bx'=1201201210 &\rightarrow \bx''=1201\underline{1}01210,\\
        \phi(\bx') = 120,0000012 &\rightarrow \phi(\bx'') = 120,0\underline{2}00\underline{1}12.
    \end{split}
\end{equation*}
\replaced{As seen in the example,}{Based on the example,} \replaced{a}{one} noisy duplication \added{of length $k$ ($k$-ND)} can be regarded as \replaced{an exact}{one} $k$-TD followed by \replaced{a}{one} substitution. Given a word $\bx\in \Sigma^{*}_{q}$, \deleted{the descendants of one $k$-TD  error satisfy $\bx'=T_{i,k}(\bx)$, } the \replaced{tandem duplication results in $\bx'=T_{i,k}(\bx)$ and the following substitution results in}{one noisy-duplication can be expressed as} $\bx''=T_{i,k}(\bx)+a \be_{j}$, where $(i+k+1)\leq j \leq (i+2k)$, $a\in \Sigma_{q}^+$, and $\be_j$ represents a unit\deleted{ary} vector with $1$ in the $j$-th entry and $0$ elsewhere. \replaced{Note that the first $k$ elements are not affected by exact or noisy duplications and $\hat\phi(\bx)=\hat\phi(\bx')=\hat\phi(\bx'')$. Hence, we focus on changes in $\bar\phi(\cdot)$.}{Since the $\hat \phi(\bx')$ always stays the same, we focus on the change of $\bar \phi(\bx')$.} \replaced{The}{One} substitution \replaced{changes}{will change} at most two symbols of $\bar \phi(\bx')$ \replaced{and}{. The substitution} can be expressed as
 \begin{equation}\label{eq:noisy_duplication}
   \deleted{\hat \phi(\bx'')=\hat \phi(\bx'), }\bar \phi(\bx'')=\bar \phi(\bx')+a \bm{\epsilon}_{j},
\end{equation}
where \replaced{$\bepsilon_j=\be_{j-k}-\be_{j}$ if $(k+1)\leq j \leq (|\bx'|-k)$ and  $\bm{\epsilon}_{j}=\be_{j-k}$ if $(|\bx'|-k+1)\leq j \leq |\bx'|$.}{the $\bm{\epsilon}^{j}$satisfies: if $(k+1)\leq j \leq (|\bx'|-k)$, $\bm{\epsilon}_{j}=\be_{j-k}-\be_{j}$; if $(|\bx'|-k+1)\leq j \leq |\bx'|$, $\bm{\epsilon}_{j}=\be_{j-k}$.} \added{We refer to $\bx''$ as a \emph{$k$-ND descendant of $\bx$.}}

Since \deleted{\replaced{the}{both the $k$-TD and the substitution errors in}}noisy duplications may occur at any position, the word $\bx$ can generate many descendants through noisy duplication errors. Let $D_{k}^{t(p)}(\bx)$ denote the \emph{descendant cone} of $\bx$ \replaced{obtained after $t$ duplications, $p$ of which are noisy}{ by $t$ $k$-TDs and $p$ substitution errors}, where $t\geq p$\deleted{, i.e. the set of sequences that can be obtained from $\bx$ with an exact or a noisy-duplication (with one substitution)}. Furthermore, the descendant cone with many exact $k$-TDs and at most $P$ \added{noisy duplications, i.e., at most $P$} substitution errors\added{,} can be expressed as
 \begin{equation}\label{eq:noisy_dup_des_cone}
   D_{k}^{*(\leq P)}(\bx)=\bigcup_{p=0}^{p=P}\bigcup_{t=p}^\infty D_k^{t(p)}(\bx).
\end{equation}
In this paper, we limit our attention to $P=1$.

\deleted{Based on the analysis above, the descendant cone $D_{k}^{*(\leq 1)}(\bx)$ contains descendants of $\bx$ with many $k$-TDs and at most one substitution error. Based on \eqref{eq:k_decre_deriv} and the example~\eqref{eq:ex1_k_TD_decre}, one $k$-TD error insert a run of $0^{k}$ to the string $\bar\phi(\bx)$ in the $k$-discrete-derivative transform. }

We define a mapping operation $\mu:\Sigma_{q}^{*} \rightarrow \Sigma_{q}^{*}$ by removing all runs of $0^{k}$ in $\bz\in \Sigma_{q}^{*}$. More specifically, consider a string $\bz$ as
\[\bz=0^{m_0}w_10^{m_1}\cdots w_{t}0^{m_{t+1}},\]
where $t=\wt(\bz)$, $w_1, \dotsc, w_{t}\in \Sigma_{q}^+$, and $m_{0}, \dotsc, m_{t+1}$ are non-negative integers. The mapping $\mu(\bz)$ \replaced{is}{can be} defined as
\[\mu(\bz)=0^{m_0\bmod k}w_10^{m_1\bmod k}\cdots w_{t}0^{m_{t+1}\bmod k}.\]
Also, $\rl(m)$ denotes the set of strings of length $m$ containing no $0^{k}$. In other words, $\rl(m)=\{\bz\in \Sigma_{q}^{m}|\mu(\bz)=\bz\}$.

According to \cite{jain2017duplication}, given a word $\bx\in \Sigma_{q}^{*}$, after many (even infinite) $k$-TD errors, the string $(\hat\phi(\bx),\mu(\bar\phi(\bx)))$ stays the same. To make use of this property, define the \emph{duplication root} $\rt(\bx)$ as the string obtained from $\bx$ after all copies of length $k$ are removed. Note that we then have
 \begin{equation}\label{eq:drt}
   \phi(\rt(\bx))=(\hat\phi(\bx), \mu(\bar\phi(\bx))).
\end{equation}
 If $\rt(\bx)=\bx$, we call the word $\bx$ irreducible. The set of all irreducible words of length $n$ can be written as $\ir(n)=\{\bx\in \Sigma_{q}^{n}|\rt(\bx)=\bx\}$. In other words, an irreducible word $\bx\in \Sigma_{q}^{n}$ satisfies $\bar \phi(\bx) \in \rl(n-k)$.

For a word $\bz\in \Sigma_{q}^{*}$, we define its \emph{indicator} $\Gamma(\bz):\Sigma_{q}^{*}\rightarrow \Sigma_{2}^{*}$ as \added{$\Gamma(\bz)=\Gamma_1(\bz)\dotsm\Gamma_{|\bz|}(\bz)$, where}
\begin{equation}\label{eq:weig_indicating_map}
    \Gamma_i(\bz)=
    \begin{cases}
          1 , & \text{if } z_i\neq 0,  \\
          0, & \text{otherwise.}
    \end{cases}
    \quad i =1,\dotsc,|\bz|.
\end{equation}

Based on \eqref{eq:noisy_duplication}, \replaced{the substitution in a noisy duplication alters two symbols in $\bar\phi(\bx')$ at distance $k$}{the noise will be added to two symbols in $\bar \phi(\bx)$ at distance $k$}. For the purpose of error correction, it will be helpful to rearrange the symbols into $k$ strings such that the two symbols affected by \replaced{the }{a }substitution appear next to each other in one of the strings.
More precisely, for $j\in[k]$, we define a \emph{splitting} operation that extracts entries whose position is equal to $j$ modulo $k$. That is, for $\bu\in\Sigma_q^n$ and $j\in[k]$, define $\ins_k(\bu,j)=\bu_j = (u_{j1}, u_{j2}, \dotsc,u_{j,\left\lfloor\frac{n-j}{k}\right\rfloor+1}) 
$ such that
\begin{equation}\label{eq:split_subj}
  u_{ji} = u_{j+(i-1)k}, \quad 1 \le i \le \left\lfloor\frac{n-j}{k}\right\rfloor+1.
\end{equation}
For $\bu\in\Sigma^n_{q}$, we then define the \emph{interleaving} operation $\inl:\Sigma_q^n\to\Sigma_q^n$ as the concatenation of $\ins_k(\bu,j), j\in[k]$,
\[
\inl(\bu)=\ins_k(\bu,1)\dotsm\ins_k(\bu,k)
.\]

\begin{example}\label{ex:SpIL}
Given an alphabet $\Sigma_{3}=\{0,1,2\}$, $k=3$, and $\bu'=\bar\phi(\bx')=2\bcomment{ 2}\rcomment{ 1}2\bcomment{ 0}\rcomment{ 0}0\bcomment{ 1}\rcomment{ 2}$,where symbols at the same position modulo $k$ have the same color, after \deleted{one} splitting \deleted{of} $\bu\added{'}$, we obtain
\begin{align*}
    \bu_1'=&\ins_3(\bu',1)=220,\\
    \bu_2'=&\ins_3(\bu',2)=\bcomment{ 201},\\
    \bu_3'=&\ins_3(\bu',3)=\rcomment{ 102},\\
    \inl(\bu')=&\bu_1'\bu_2'\bu_3'=220\bcomment{201} \rcomment{102}.
\end{align*}
Based on \eqref{eq:noisy_duplication}, after one substitution error, we may obtain $\bu''=\bar\phi(\bx'')=2\bcomment{ 2}\rcomment{ 1}2\bcomment{ 0}\rcomment{ \underline{1}}0\bcomment{ 1}\rcomment{\underline{1}}$, where symbols affected by the substitution error are underlined. \replaced{We then have}{Similarly, the interleaving string}
\begin{align*}
    \bu_1''=&\ins_3(\bu'',2)=201,\\
    \bu_2''=&\ins_3(\bu'',1)=\bcomment{220},\\
    \bu_3''=&\ins_3(\bu'',3)=\rcomment{ 1\underline{11}},\\
    \inl(\bu'')=&\bu''_1\bu''_2\bu''_3=220\bcomment{ 201}\rcomment{ 1\underline{11}}.
\end{align*}
\deleted{$\inl(\bu'')=\ins(\bu'',1)\ins(\bu'',2)\ins(\bu'',3)=2202011\underline{11}$.} \replaced{We observe that the error is restricted to $\bu_3''$ and that the two symbols changed by the substitution error are adjacent in $\inl(\bu'')$, while they are not so in $\bu''$.}{$\inl(\bu'')$, the interleaving can adjust the two changed symbols with distance $k$ to be two adjacent symbols.}
\end{example}

Given a word $\bz\in \Sigma_{q}^{n}$, we define the \emph{cumulative-sum} operation $\cusum:\Sigma_{q}^{n} \rightarrow \Sigma_{q}^{n}$, as $\br=\cusum(\bz)$, where \begin{equation}\label{eq:cumulative_sum}
   r_i=\sum^{i}_{t=1}z_t \bmod q,\quad i=1,\dotsc, n.
\end{equation}

\replaced{We further}{Apart from inter-splitting operation, we} define the \emph{odd \replaced{subsequence}{string}} \replaced{$\Od(\bz)$}{$\bar\bz_1$} and the \emph{even \replaced{subsequence}{string}} \replaced{$\Even(\bz)$}{$\bar\bz_2$} of a word $\bz\in \Sigma_{q}^{*}$ as two sequences containing symbols in the odd and even positions\added{,} respectively. More precisely, \replaced{$\Od(\bz)=\ins_2(\bz,1)$ and $\Even(\bz)=\ins_2(\bz,2)$.}{for $\bz\in\Sigma_{q}^{n}$ and $j\in[2]$, we define the odd string $\bar\bz_1=(\bar z_{1,i})_i=\OdEv(\bz,1)$ and even string $\bar\bz_2=(\bar z_{2,i})_i=\OdEv(\bz,2)$ such that}

Our results will rely on codes that can correct a single insertion or deletion\replaced{. We}{, we} thus recall the Varshamov-Tenengolts codes \cite{sloane2000single,tenengolts1984nonbinary}\added{, which are binary codes capable of correcting a single insertion or deletion (indel)}.

\begin{construction}[\cite{sloane2000single}]\label{Con:vt_code}
Given \replaced{integers}{an integer} $m\geq 1$ and $0\leq \alpha\leq (m-1)$, the binary Varshamov-Tenengolts (VT) code $C_{VT}(\alpha,m)$ is given as
\begin{equation}
   C_{VT}(\alpha,m)=\{\bz\in \Sigma_{2}^{\le m-1}| \sum_{i=1}^{|\bz|}iz_{i}=\alpha \bmod m\}.
\end{equation}
\end{construction}
We note that there is a minor difference between this construction and the original VT code. Namely, we allow strings of length at most $m-1$ rather than exactly $m-1$. If the length of the stored word is known, it follows from the proof of the VT code that the code in the construction above can correct a singel indel.

Compared to the binary indel-correcting code, correcting indels in non-binary sequences is more challenging. \replaced{We will use Tenengolts' $q$-ary single-indel-correcting code~\cite{tenengolts1984nonbinary}, which relies on the mapping $\zeta:\Sigma_{q}^{*}\to \Sigma_{2}^{*}$, where}{Based on \cite{tenengolts1984nonbinary}, we first define one \emph{unique-binary mapping} $\zeta:\Sigma_{q}^{*}\to \Sigma_{2}^{*}$ by comparing two symbols in adjacent positions. More specifically,} the $i$-th position of $\zeta(\bz)$ is
\begin{equation}\label{eq:Unique_binary}
  \zeta_i(\bz)=
    \begin{cases}
          1 , & \text{ if } z_i \geq z_{i-1},   \\
          0, & \text{ if } z_i < z_{i-1}.
    \end{cases}
    \qquad i=2,3,\dotsc, |\bz|.
\end{equation}
\replaced{with}{where} $\zeta_1(\bz)=1$.

\begin{construction}[\cite{tenengolts1984nonbinary}]\label{con:Tq_Code}
For integers $m\geq 1$, $0\leq \alpha\leq (q-1)$ and $0\leq \beta\leq (m-1)$, Tenengolts' $q$-ary single indel-correcting code $C_{Tq}(\alpha,\beta, m)$ over $\Sigma_{q}^{\leq m}$ is given as
 \begin{equation}
 \begin{split}
       C_{Tq}(\alpha,&\beta,m)=\bigg\{ \bz\in \Sigma_{q}^{ \leq m}\bigg\vert \sum_{j=1}^{|\bz|}z_{j}=\alpha \bmod q,\\& \sum_{i=1}^{|\bz|}(i-1)\zeta_i(\bz)=\beta \bmod m\bigg\}.
 \end{split}
\end{equation}
\end{construction}
Again, we allow codewords of length at most $m$, rather than exactly $m$ as was the case in Tenengolts' original construction. If the length of the stored codeword is known, it follows from the proofs of the VT code that we can recover the binary sequence $\zeta_i(\bz)$ with $|\bz|\le m$. Then the code in the construction above can correct a single indel in $\bz$.

 Given a family of codes $\cC=\{C_n\}_n$, where $C_n\subseteq \Sigma_{q}^{n}$, the \emph{code rate} is defined as
\begin{equation}\label{eq:rate-def}
    R_n(\cC)=\frac{1}{n} \log_q |C_n|,
\end{equation}
where $|C|$ denotes the size of the code $C$. Furthermore, the \emph{asymptotic rate}
is defined as
\begin{equation}\label{eq:cap-def}
   R(\cC)=\limsup_{n\to \infty}R_n(\cC).
\end{equation}
If the meaning is clear from the context, we may refer to both the family and individual codes  as $C$ and write $R_n(C),R(C)$ to refer to the rates for a given construction.

\section{Noisy duplication channels}\label{sec:descands_noi_channel}
To enable designing error-correcting codes, in this section, we study the relation between the input and output sequences in \emph{noisy duplication channels}. As before, we consider channels with many (possibly infinite) exact duplications and at most one noisy duplication in which one of the copied symbols is altered.


If a code $C\in \Sigma_{q}^{n}$ corrects many $k$-TD and one $k$-ND errors, then for any two distinct codewords $\bc_{1},\bc_{2}\in C$, we have
\begin{equation}\label{eq:code_drt_nointersection}
      D_k^{*(\le1)}(\bc_1)\cap D_k^{*(\leq1)}(\bc_2) = \varnothing,
\end{equation}
where $D_k^{*(\le1)}(\cdot)$ is defined in \eqref{eq:noisy_dup_des_cone}.
This can be shown to be equivalent to
\begin{equation}\label{eq:code_drt}
\begin{split}
 \rt(\bc_2)&\neq\rt(\bc_1),\\
 \rt(D_k^{*(\leq 1)}(\bc_1))&\cap\rt(D_k^{*(\leq 1)}(\bc_2))= \varnothing.
\end{split}
\end{equation}

Since $k$-TDs do not alter the root of the sequence, $\rt(\bc_2)\neq\rt(\bc_1)$ ensures that $k$-TD errors can be corrected. Noisy tandem duplications however alter the roots. In fact, they may produce sequences with roots whose lengths are different from the roots of the stored sequences. Since the codewords have distinct roots, it suffices to recover the root of the retrieved word to correct any errors. We will restrict our constructions to codes whose codewords are irreducible, and thus are their own roots. While this is not necessary, it will simplify the code construction, as we will show, and does not incur a large penalty in terms of the size of the code. 

For noisy duplication channels, given a codeword $\bx\in \Sigma_{q}^{n}$, the generation of descendants $\bx''\in D_k^{*(\leq 1)}(\bx)$ includes three different cases: only $k$-TDs; $k$-TDs followed by one $k$-ND; and $k$-TDs, followed by a $k$-ND, followed by more $k$-TDs. Since the root is not affected by the $k$-TDs, to study $\rt(D_k^{*(\leq 1)}(\bx))$, we only need to consider the second case, i.e., 
we focus on descendants $\bx''$ immediately after the noisy duplication.

Given an irreducible string $\bx \in \Sigma_{q}^{n}$ with $n>2k$, our goal is to characterize $\rt(D_k^{*(\leq 1)}(\bx))$. Based on \eqref{eq:k_decre_deriv}, we have
\begin{equation}
    \phi(\bx)=(\hat \phi(\bx), \bar \phi(\bx))=(\by,\bz),
\end{equation}
where $\by=\hat\phi(\bx)\in \Sigma_{q}^{k}$ and $\bz=\bar\phi(\bx)\in \Sigma_{q}^{n-k}$. Since $\bx$ is an irreducible string, the string $\bz$ contains no runs of $0^{k}$, i.e. $\bz=\mu(\bz)$.

After many $k$-TDs and one $k$-ND, we have a descendant $\bx''\in D_k^{*(\leq 1)}(\bx) $. Since the substitution only occurs in the copy, the first $k$ symbols always stay the same. Thus $\bx''$ satisfies
\begin{equation}
    \phi(\bx'')=(\hat \phi(\bx''), \bar \phi(\bx''))=(\hat \phi(\bx), \bar \phi(\bx''))=(\by,\bz'').
\end{equation}

Based on \eqref{eq:drt}, it suffices to study the problem in the transform domain, i.e., we want to obtain all possible $(\by,\mu(\bz''))$ derived from $(\by,\mu(\bz))$. 
Our code constructions in the next section will also rely on certain sequences derived from $\mu(\bz)$. The next theorem characterizes how these sequences can be altered by $k$-TDs and one $k$-ND. The theorem relies on the indicator map $\Gamma$, defined in~\eqref{eq:weig_indicating_map}, and on the splitting operation defined in~\eqref{eq:split_subj}.

\begin{theorem}\label{Theo:descendant_changes}
Let $\bx\in\Sigma_q^n$ and let $\bx''\in D_k^{*(\le 1)}(\bx)$ be a descendent of $\bx$ (produced by passing through the noisy duplication channel). Furthermore, let
\begin{align*}
    \bz &= \bar\phi(\bx), &    \bmu & = \mu(\bz),\\
    \bmu_j &= \ins_k(\bmu,j),&     \bs_j &= \Gamma(\bmu_j),
\end{align*}
Then
we define $\bz'',\bmu'',\bmu_j'',\bs_j''$, similarly, based on $\bx''$. The differences between sequences defined based on $\bx$ and $\bx''$ are given in Table~\ref{Table:descendant_change_case} and Table~\ref{Table:descendant_change_case_tail}.

\begin{table*}[]
    \centering
    \caption{The changes in $\bmu_j$ and $\bs_j$, $j\in [k]$ as a result of exact and noisy duplications, when the position of the substitution in $\bx''$ satisfies $k< p\leq (|\bx''|-k)$. Here $a,b,c\in\Sigma_q$, $d\in\Sigma_2$, $ \bar a = -a$, and $a,b\neq 0$. Furthermore, $\Lambda\to \bu$ and $\bu\to \Lambda$ represent insertion and deletion of the string $\bu$, respectively. Rows marked by $(*)$ indicate that this type of error occurs for at most one value of $j\in[k]$. The marking $(\$)$ is related to the error-correction strategy discussed in Section~\ref{sec:correct_code}
    .}
    \label{Table:descendant_change_case}
    \begin{tabular}{p{1.3cm}|p{3.3 cm}|p{3cm}|p{4cm}}
    \hline
        $|\bmu''|-|\bmu|$ & $\bmu\to\bmu''$
        & $\bmu_j\to\bmu_j''$   & $\bs_j\to\bs_j''$\\
        \hline\hline
        $+2k$ & insert $0^{j-1}a0^{k-j}$ and $0^{t-1}(0-a)0^{k-t}$& $\Lambda\to a\bar a$ \hfill $(*)$\newline $\Lambda\to00$ \hfill$(\$)$ \newline $c\to 0c0$ \hfill$(\$)$& $\Lambda\to 11$\newline $\Lambda\to00$ \newline $d\to 0d0$\\ \hline\hline
        \multirow{2}{*}{$+k$} & {insert $0^{j-1}a0^{k-j}$ and substitute $b_i\rightarrow (b_i-a)$ } & $c\to a(c-a),c\neq a$\hfill $(*)$\newline $a\to a0\ (\Lambda\to0)$ \hfill$(\$)$ \newline$\Lambda\to 0$ \hfill$(\$)$ & $0\to 11, 1\to11$\newline$1\to10\ (\Lambda\to 0)$\newline $\Lambda\to 0$\\
            \cline{2-4}
            & {substitute $0\rightarrow a$ and insert $0^{t-1}(0-a)0^{k-t}$}  &  $0\to a\bar a$\hfill $(*)$\newline $\Lambda\to 0$ \hfill$(\$)$&  $0\to 11$\newline $\Lambda\to 0$ \\
        \hline\hline
        \multirow{2}{*}{0} &  {insert $0^{j-1}a0^{k-j}$ and delete $0^{t-1}a0^{k-t}$  with $a$ at the same position}  & $b0\to 0b$ \hfill$(\$)$\newline stay same & $10\to 01$\newline stay same \\
            \cline{2-4}
            & {substitute $0\rightarrow a$ and $b_i\rightarrow (b_i-a)$ with distance $k$}  & $0c\to a(c-a)$\hfill $(*,\$)$\newline stay same& $00\to11,01\to11,01\to10$\newline stay same \\
        \hline\hline
        $-k$ & substitute $0\rightarrow a$ and delete $0^{t-1}a0^{k-t}$ & $0\to \Lambda$ \hfill$(\$)$&$0\to \Lambda$  \\
    \hline
    \end{tabular}
\end{table*}

\begin{table*}[]
    \centering
    \caption{The changes in $\bmu_j$ and $\bs_j$, $j\in [k]$ as a result of exact and noisy duplication, when the position of the substitution in $\bx''$ satisfies $ (|\bx''|-k)<p\leq |\bx''|$. The notation is the same as that of Table~\ref{Table:descendant_change_case}.}      \label{Table:descendant_change_case_tail}
    \begin{tabular}{p{1.3cm}|p{3.3cm}|p{3cm}|p{4cm}}
    \hline
        $|\bmu''|-|\bmu|$ & $\bmu\to\bmu''$
        & $\bmu_j\to\bmu_j''$   & $\bs_j\to\bs_j''$ \\
        \hline\hline

        $+k$ & {insert $0^{j-1}a0^{k-j}$ } & $\Lambda\to a$\hfill $(*)$\newline$\Lambda\to 0$ \hfill$(\$)$& $\Lambda\to 1$\newline $\Lambda\to 0$\\
        \hline\hline
        $0$ & {substitute $0\rightarrow a$}  & $0\to a$\hfill $(*,\$)$\newline stay same& $0\to1$\newline stay same \\

    \hline
    \end{tabular}
\end{table*}
\end{theorem}
The theorem is proved in the appendix.
To illustrate the theorem, we provide an example (in the transform domain). 

\begin{example}\label{examp:channel_descendants}
Consider $\Sigma_3=\{0,1,2\}$, $k=3$, and $\bmu=\mu(\bz)=\bz=120102002120$. Suppose that after several $k$-TDs, the descendant is $\bz'=0^31 0^3200^31020020^610^320$. Next a $k$-ND may insert a substring $0^3$ (marked red below) and alter one or two symbols (underlined). Depending on the positions of the duplication and substitution, the following cases are possible:
\begin{itemize}
     \item If $\bz''=0^3\rcomment{0\underline{2}0}1 \underline{1}00200^31020020^610^320$, then $\bmu''=\mu(\bz'')=020110020102002120$ and $|\bmu''|-|\bmu|=2k$, as in the $1$st row of Table~\ref{Table:descendant_change_case}.
    \item If $\bz''=0^31 0^3200^31020020^610^3\rcomment{0\underline{2}0}2\underline{1}$, then $\bmu''=\mu(\bz'')=120102002102021$ and $|\bmu''|-|\bmu|=k$, as in the $2$nd row of Table~\ref{Table:descendant_change_case}.
    \item If $\bz''=0^31 0^3200^3\rcomment{00\underline{1}}10\underline{1}0020^610^320$, then $\bmu''=\mu(\bz'')=121101002120$ and $|\bmu''|=|\bmu|$, as in the $3$rd row of Table~\ref{Table:descendant_change_case}.
     \item If $\bz''=0^31 0^3200^3\rcomment{00\underline{2}}10\underline{0}0020^610^320$, then $\bmu''=\mu(\bz'')=122102120$ and $|\bmu''|-|\bmu|=-k$, as in the $4$th row of Table~\ref{Table:descendant_change_case}.
\end{itemize}
\end{example}

\deleted{We will first discuss the results given by the theorem. First,} Since the length of $\bmu$ can change by $-k$, $0$, $k$, or $2k$, the noisy duplication may manifest as deletions, insertions, or substitutions in $\bmu$. Furthermore, the complex error patterns in $\bmu$ are simplified when we consider $\bmu_j, j\in [k]$. The errors marked by $(*)$ occur for at most one value of $j$. These correspond to positions affected by the substitution. (Rows marked by $(\$)$ relate to our error-correction strategy and are discussed in the next section.

We note that for correcting any number of exact duplications and $t$ noisy duplications, each containing a single substitution, a description of the channel can be obtained based on Tables~\ref{Table:descendant_change_case} and \ref{Table:descendant_change_case_tail}. This is because the tables describe the effect of a sequence of many exact duplications and one noisy duplication on the root of the sequence (and its derived subsequences) and because a sequence of errors containing $t$ noisy duplications can be divided into $t$ parts, each consisting of a number of exact duplications and a single noisy duplication. In particular, the length of the root may change by $-2k,-k,0,k,2k,3k,$ or $4k$ for two noisy duplications. If each noisy duplication contains more than one substitution, however, characterizing the channel becomes more challenging as the number of possible cases grows.

Now that we have determined all changes from $(\by,\bmu)$ to $(\by,\bmu'')$ resulting from passing through the noisy duplication channel, we consider the code design to correct many exact $k$-TDs and at most one noisy duplication in the next section.


\section{Error-correcting codes\\ for noisy duplication channels}\label{sec:correct_code}
Recall from Section~\ref{sec:descands_noi_channel} that we are interested in constructing a code $C\subseteq \ir(n) \cap \Sigma_{q}^{n}$ that can correct many exact $k$-TDs and at most one noisy duplication. Based on \eqref{eq:code_drt}, for any code that corrects $k$-TDs, two distinct codewords must have distinct roots. Thus, for a stored codeword $\bx$ and the retrieved word $\bx''$, if we can recover the duplication root $\rt(\bx)$ of $\bx$ from $\bx''$, we can recover the codeword $\bx$. But we have made a further simplifying assumption that $C\subseteq \ir(n)$ and thus $\bx=\rt(\bx)$.

As shown in Theorem~\ref{Theo:descendant_changes}, duplication errors manifest in various ways in $\rt(\bx'')$ and its counterpart in the $\mu$-transform domain $\mu(\bar\phi(\bx''))$. Hence, for error correction, we utilize several sequences derived from  $\bx$, including $\bmu_j$ and $\bs_j$, $j\in[k]$, as defined in Theorem~\ref{Theo:descendant_changes}. Furthermore, we define $\br=\cusum(\inl(\bmu))$ and $\br''=\cusum(\inl(\bmu''))$. We note that $\br$ (similarly $\br''$) can be directly found by rearranging the elements $x_{k+1}\dotsm x_n$.

\begin{figure*}
    \begin{center}
        \begin{tikzpicture}[thick, scale=0.8, vnd/.style={shape=circle,draw,inner sep=0,outer sep=0, minimum width=1cm, scale=0.8},
        invertibleedge/.style={->, line width=.3mm}]
            \node (x) [vnd] {$\bx$};
            \node (z) [above right = of x, vnd] {$\bz$};
            \draw [invertibleedge] (x) -- (z) node [midway, above, rotate=45] {$\bar\phi$};
            \node (drtx) [below right = of x, vnd] {$\rt(\bx)$};
            \draw [->,dashed] (x) -- (drtx) node [midway, below, rotate=-45] {$\rt$};
            \node (mu) [below right = of z, vnd] {$\bmu$};
            \draw [->,dashed] (z) -- (mu) node [midway, above, rotate=-45] {$\mu$};
            \draw [invertibleedge] (drtx) -- (mu) node [midway, below, rotate=45] {$\bar\phi$};
            \node (muj) [above right = of mu, vnd] {$\bmu_j$};
            \draw [invertibleedge] (mu) -- (muj) node [midway, above, rotate=45] {$\ins_k$};
            \node (sj) [below right = of mu, vnd] {$\bs_j$};
            \draw [->,dashed] (muj) -- (sj) node [midway, above, rotate = -90] {$\Gamma$};
            \node (Imu) [below right = of muj, vnd] {$\inl(\bmu)$};
            \draw [invertibleedge] (muj) -- (Imu) node [midway, above, rotate=-45] {Concat.};
            \node (r) [right = of Imu,vnd] {$\br$};
            \draw [invertibleedge] (Imu) -- (r) node [midway, above] {$\cusum$};
        \end{tikzpicture}
    \caption{The various mapping used in the paper. ``Concat.'' stands for concatenation. Solid edges indicate invertible mappings, where we have assumed $x_1\dotsm x_k$ is known, since these symbols are not affected by the channel. The mapping $\mu$ is generally non-invertible, but in our constructions, since we assume $\bx$ is irreducible, if we recover $\bmu=\mu(\bx)$, we can recover $\bx$. } \label{Fig:sequences_code_cons}
    \end{center}
\end{figure*}
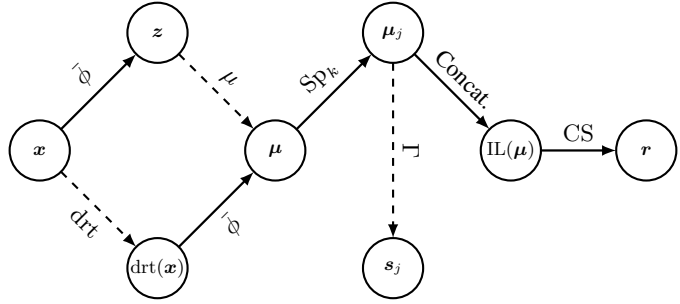

The relationship between these mappings is illustrated in Figure~\ref{Fig:sequences_code_cons}. In the figure, solid edges represent invertible mappings. Since $\bx$ is irreducible, the stored codeword can be recovered if any of $\bmu$, $(\bmu_j)_{j\in[k]}$, $\inl(\bmu)$ or $\br$ are recovered (note that $x_1\dotsm x_k$ are not affected by errors). We use these mappings to simplify and correct different error
patterns described by Theorem~\ref{Theo:descendant_changes} in an efficient manner.

The motivation behind defining $\bmu_j$, $j\in[k]$, is to convert insertions and deletions of blocks of length $k$ into simpler errors involving one or two symbols. Some of the errors, marked by $(\$)$ in Tables~\ref{Table:descendant_change_case} and~\ref{Table:descendant_change_case_tail}, involve $0$s, which appear in the same positions in $\bs_j$ and $\bmu_j$. Correcting these errors in $\bs_j$ is more efficient since it will rely on binary codes rather than $q$-ary codes. We will first correct these errors in $\bs_j$ and then correct the corresponding $\bmu_j$. Finally, the cumulative-sum mapping $\cusum$ turns errors marked by $(*)$, e.g., $\Lambda\to a\bar a$ into a single $q$-ary insertion or substitution. Importantly, in each case there is only one such error. So if other errors are corrected, we can concatenate $\bmu_j$, $j\in[k]$, and then correct the single occurrence of this error.

We will construct an error-correcting code that will allow us to recover $\bmu$ from $\bmu''$. As discussed, for certain errors occurring in $\bmu_j$, specifically those marked by $(\$)$ in Tables~\ref{Table:descendant_change_case} and~\ref{Table:descendant_change_case_tail}, we may do so by correcting errors in $\bs_j$, via Construction~\ref{cons:code_binary_errors} below.

The indicator vectors $(\bs_{1},\dotsc, \bs_{k})$ are subject to several error patterns: insertion of $11$; insertion of two $0$s with distance at most $2$; indel of $1$ or $0$; swaps of two adjacent elements; and substitution of one or two 0s with one or two 1s. 
The following code can correct a single occurrence of one of these errors, as shown in the next theorem. A slightly modified version of this code is used for the noisy duplication channel.

\begin{construction}\label{cons:code_binary_errors}
Given integers $0\leq a\leq 2(n+1)$, $0\leq b\leq 4$, and $0\leq c\leq 2n$, we construct the code $C_{(a,b,c)}$ as
\begin{align}
      C_{(a,b,c)} =\{\bu \in \Sigma^n_2|&\label{eq:binary_VT} \bu \in C_{VT}(a, 2n+3),\\
                    &\label{eq:binary_weight_change} \sum^{n}_{i=1}  u_i=b \bmod 5, \\
                    &\label{eq:binary_transition} \sum^{n}_{i=1} i \left( \sum^{j=i}_{j=1} u_j\right)=c \bmod (2n+1)\},
\end{align}
where $n=|\bu|$.
\end{construction}

\begin{theorem}\label{Theo:Binary_code}
The code $C_{(a,b,c)}$ can correct a single occurrence of any of the following errors (without a priori knowledge of the type of error):
\begin{itemize}
    \item an insertion, deletion, or substitution,
    \item a substitution of two adjacent bits,
    \item a substitution of one bit by two adjacent bits,
    \item an insertion of two bits of the form $\Lambda\to 11$, $\Lambda\to 00$, or $1\to010$.
    \end{itemize}
These error patterns include all those shown in the $\bs_j$ column of Tables~\ref{Table:descendant_change_case} and~\ref{Table:descendant_change_case_tail}.
\end{theorem}

The theorem is proved in the appendix.

Since $(\bs_{1},\dotsc, \bs_{k})$ are weight indicators of $(\bmu_{1},\dotsc, \bmu_{k})$, the $0$s in $(\bs_{1},\dotsc, \bs_{k})$ and $(\bmu_{1},\dotsc, \bmu_{k})$ coincide. However, if a 1 is deleted from a run of 1s in $\bs_j$, we will not be able to identify which symbol is deleted from $\bmu_j$. This means that after recovering $\bs_j$ from $\bs_j''$ we can recover $\bmu_j$ only in certain cases, specifically, those marked 
by $(\$)$ in Table~\ref{Table:descendant_change_case} and Table~\ref{Table:descendant_change_case_tail}.
Interestingly, the errors not corrected by recovering $\bs_j,j\in[k]$ are marked by $(*)$, indicating that they occur only for a single value of $j$. Hence, to correct these errors, we apply the code constraints to the concatenation of $\bmu_j,j\in[k],$ rather than to each $\bmu_j$ separately.

\begin{construction}\label{con:noisy_duplication_code}
Define $C_{nd}\subseteq\Sigma_{q}^{n}$ as
  \begin{align}
     C_{nd} =\{& \bx\in \ir(n)\cap \Sigma_{q}^{n}| \bmu=\mu(\bar\phi(\bx)),\\& \bmu_j=\ins_k(\bmu,j), \bs_j=\Gamma(\bmu_j),\\
     & \label{eq:nd_VT}\bs_{j}\in C_{VT}(a_{j}, 2|\bs_{j}|+3), \\
    &\label{eq:nd_transposition} \sum^{|\bs_{j}|}_{i=1} i \left( \sum^{t=i}_{t=1} s_{jt}\right)=c_{j} \bmod (2|\bs_{j}|+1),\\
    &\label{eq:nd_weight} \sum^{k}_{j=1} \sum^{|\bs_{j}|}_{i=1} s_{ji}=b \bmod 5, \\
    & \label{eq:nd_Tq_odd} \Od(\inl(\bmu))\in C_{Tq}(\bar a_1,\bar b_1,\lceil\frac{n-k}{2}\rceil),\\
    & \label{eq:nd_Tq_even} \Even(\inl(\bmu))\in C_{Tq}(\bar a_2,\bar b_2,\lceil\frac{n-k}{2}\rceil),\\
    & \label{eq:nd_Tq_Culsum} \cusum(\inl(\bmu)) \in C_{Tq}(\bar a_3,\bar b_3,n-k),\\
    & \,\inl(\bmu) \in C_{Tq}(\bar a_4,\bar b_4,n-k)\},\label{eq:nd_Tq_concatinating}
  \end{align}
where $j, a_j, c_j,b,\bar a_i, \bar b_i$ are integers satisfying $  j\in [k]$, $0\leq a_{j}\leq 2(|\bs_{j}|+1)$,  $0\leq c_{j} \leq 2|\bs_{j}|$,  $0\leq b\leq 4$,  $0\leq \bar a_1,\bar a_2,\bar a_3, \bar a_4<q$, $0\leq \bar b_1, \bar b_2 \leq \lfloor\frac{n-k}{2}\rfloor$, and $0\leq \bar b_3,\bar b_4 < n-k$.
\end{construction}

In Construction~\ref{con:noisy_duplication_code}, the constraints \eqref{eq:nd_VT}, \eqref{eq:nd_transposition}, and \eqref{eq:nd_weight} play the same role as the code in Construction~\ref{cons:code_binary_errors}, and the constraints~\eqref{eq:nd_Tq_odd}, \eqref{eq:nd_Tq_even}, \eqref{eq:nd_Tq_Culsum}, and  \eqref{eq:nd_Tq_concatinating} can correct the error patterns of $\{\bmu_{1},\dotsc,\bmu_{k}\}$ not marked by $(\$)$ in Table~\ref{Table:descendant_change_case} and Table~\ref{Table:descendant_change_case_tail}.
The constraint~\eqref{eq:nd_VT} corrects one insertion/deletion or two insertions of $0$s or $1$s in adjacent positions over $\Sigma_{2}$. The constraint \eqref{eq:nd_transposition} corrects one transposition of $\{0,1\}$ in two adjacent positions. The constraint \eqref{eq:nd_weight} is a weight-indicating equation for $\{\bs_{1},\dotsc,\bs_{k}\}$. The constraints \eqref{eq:nd_Tq_odd}, \eqref{eq:nd_Tq_even}, \eqref{eq:nd_Tq_concatinating}, and \eqref{eq:nd_Tq_Culsum} can correct one insertion/deletion \replaced{in}{of} $\Od(\inl(\bmu))$, $\Even(\inl(\bmu))$, $\inl(\bmu)$, and $\br=\cusum(\inl(\bmu))$ over $\Sigma_{q}$, respectively.

\begin{theorem}\label{Theom:nd_error_code}
The error-correcting code $C_{nd}$ proposed in Construction~\ref{con:noisy_duplication_code} can correct infinitely many exact $k$-TD and up to one $k$-ND errors. There exists one such code with size
\begin{equation}\label{eq:code_size}
    \frac{|\ir(n)|}{5q^{4}\lceil\frac{n-k}{2}\rceil^2(4\lceil\frac{n}{k}\rceil^2-1)^{k}(n-k)^{2}}\leq|C_{nd}|\leq|\ir(n)|.
\end{equation}
\end{theorem}



The proof is given in the appendix. From~\eqref{eq:code_size}, we have
\begin{equation*}\label{eq:code_rate_define}
\begin{split}
       \frac{1}{n}\log_q|\ir(n)|-\frac{2k+4}{n}\log_qn-\frac{2k+5}{n} \leq R_n(C_{nd})\leq \frac{1}{n}\log_q|\ir(n)|.
\end{split}
\end{equation*}
Furthermore, based on 
\cite[(8)]{tang2019single}, for $q+k\ge 4$, 
$\frac{M}{2}\le |\ir(n)|\le M,$
where $M\eqdef\sum_{i=0}^{\lfloor n/k\rfloor-1} |\ir(n-ik)|$ is the size of the optimal code of length $n$ that can correct any number of exact duplications
.
Hence,
\begin{equation*}\label{eq:code_rate_define_bound}
\begin{split}
       \frac{1}{n}\log_qM-\frac{2k+4}{n}\log_qn-\frac{2k+6}{n} \leq R_n(C_{nd})\leq \frac{1}{n}\log_qM.
\end{split}
\end{equation*}
In particular, compared to the optimal code correcting only exact duplications, the redundancy is $\lesssim(2k+4)\log_qn$ symbols. Additionally both codes have the same asymptotic rate (given in~\eqref{eq:kTDrate} for large $k$), and in this sense the code proposed here is asymptotically optimal, although it is not clear whether $(2k+4)\log_qn$ is the best possible redundancy.

\begin{figure}
    \begin{center}
    \includegraphics[width=0.5\textwidth]{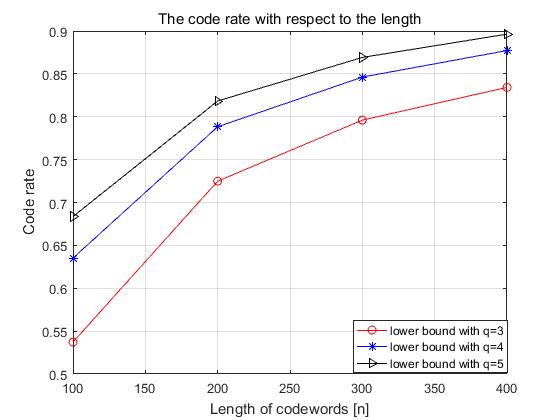}
    \caption{The lower bound of the code rate with respect to the length $n$ with the duplication length $k=3$ and alphabet size $q\in \{3,4,5\}$.}
    \label{fig:rate_nary_qary_k3_lowvbound}
    \end{center}
\end{figure}

For the alphabet size $q\in\{3,4,5\}$ and the duplication length $k=3$, Figure~\ref{fig:rate_nary_qary_k3_lowvbound} shows the lower bound of the code rate  
as the length $n$ of codewords ranges from 100 to 400, based on \eqref{eq:code_size}, \eqref{eq:rate-def} and \cite{jain2017duplication}. 

\section{Conclusion}
In this paper, we considered the problem of correcting exact and noisy duplication errors and focused on the case of many exact duplications and one noisy duplication, which suffers from a substitution. Our error-correction strategy is based on roots of sequences and splitting operations that distribute the effect of the duplication among $k$ subsequences of the root but bring symbols affected by a substitution together. In Section~\ref{sec:descands_noi_channel}, we characterized the effect of the single noisy duplication channel on the roots and their subsequences. We constructed an error-correcting code that first recovers certain binary substrings with whose help we can determine $k-1$ of the subsequences of the root, leaving a $q$-ary error in the last subsequence, which is corrected using Tenengolts' $q$-ary code. Finally, we showed that relative to the optimal code for correcting only exact duplications, the proposed construction incurs a redundancy of approximately $(2k+4)\log_q n$ symbols.

Directions for future work include correcting more than one noisy duplication while still correcting any number of exact duplications. The characterization of the channel for one noisy duplication presented in Theorem~\ref{Theo:descendant_changes}, which was based on roots and their subsequences, can be useful for describing channels allowing more noisy duplications, especially if the noise level in duplications is limited so that only a single substitution may occur in each. Then, appropriate extensions of more powerful binary deletion-correcting codes~\cite{sima_two_2018,gabrys_codes_2019,sima_optimal_2019} could be useful for correcting errors similar to those described in Theorem~\ref{Theo:Binary_code}. Codes over the full alphabet could then be used to correct the smaller number of $q$-ary errors that remain. If the number of substitutions in each noisy duplication is larger, while Theorem~\ref{Theo:descendant_changes} will still provide insights into the problem, characterizing the channel becomes more challenging. If the noise level in each duplication is so high that most of the duplicated symbols are substituted, the benefit of first correcting binary errors diminishes as there will be more errors of the type $\Lambda\to a\bar a$ than $\Lambda\to 00$, say. Dividing the root into subsequences, however, will still be useful for distributing the errors among multiple sequences. Finally, if the total number of duplications is limited, restricting codewords to irreducible sequences will be inefficient, although a combination of irreducible sequences and codes in the $\ell_1$-metric have been found to be of use in the related problem of correcting only exact duplications~\cite{jain2017capacity}. Another possible direction is the use of constrained codes, similar to~\cite{tang2019single}, to decrease the variety of the error types that may be encountered, thereby simplifying code construction. We note however that such an approach will likely incur a rate penalty.

\bibliographystyle{IEEEtranS}
\bibliography{references}

\begin{appendices}
\section{Proof of Theorem~\ref{Theo:descendant_changes}}
\begin{IEEEproof}\label{proof:descendant_changes_case}
In a noisy duplication channel with many exact $k$-TDs and at most one $k$-ND, given a string $\bx\in \Sigma^{*}_{q}$, let $\phi(\bx)=(\by,\bz)$ with $\by=\hat \phi(\bx)\in \Sigma^{k}_{q}$ and $\bz=\bar\phi(\bx)\in \Sigma^{*}_{q}$. Since the $k$-TDs do not change the duplication root $\rt(\bx)$, we focus our attention to the substitution that will change the duplication root. After many exact $k$-TDs, we obtain $\bx'\in D_k^{\geq 1(0)}(\bx)$, a descendant of $\bx$. After the substitution error, we have $\bx''
\in D_k^{\geq1(1)}(\bx)$. Since the following $k$-TD errors do not change the duplication root $\rt(\bx'')$, we focus on the descendants $\bx'$ and $\bx''$.

Let $\phi(\bx')=(\by,\bz')$ and $\phi(\bx'')=(\by,\bz'')$. In the transform domain, the string $\bz'$ can be expressed as
  \[ \bz'= \bu a_{1}a_{2} \cdots a_{i}\cdots  a_{k} b_{1} b_{2}\cdots b_{i}\cdots b_{k}\bv.\]
where $\bu,\bv \in \Sigma^{*}_{q}$ and $a_{i},b_{i} \in \Sigma_{q}, i\in[k]$. Let the length of the run of $0$s on the left side of $a_{i}$ be $m_{1}$ and on the right side of $a_{i}$ be $m_{2}$ (ending at $b_i$ and excluding $a_{i},b_i$), i.e., the substring $c0^{m_1}a_i0^{m_2}d$ with $a,b\in\Sigma_{q}^{+}$. Similarly, we define $m_{3}$ and $m_{4}$ as the length of the run of $0$s on the left side and right side of $b_{i}$, starting from $a_i$ and excluding $a_i,b_i$.
Based on~\eqref{eq:noisy_duplication}, if the substitution position $p$ satisfies $k<p\leq (|\bx'|-k)$, the substitution changes two symbols; if $(|\bx'|-k)<p\leq |\bx'|$, the substitution changes one symbol.

First, we consider the substitution position satisfying $k<p\leq (|\bx'|-k)$ such that two symbols of $\bz'$ changes. The $2$ symbols in $\bz'$ have a distance of $k$. After the substitution, we have
  \[ \bz''= \bu a_{1}a_{2} \cdots (a_{i}+a)\cdots a_{k} b_{1} b_{2}\cdots(b_{i}-a)\cdots b_{k}\bv,\]
where $a\in \Sigma_{q}^{+}$. Based on~\eqref{eq:ex1_k_TD_decre}, since the substitution only occurs in the copy of a $k$-TD, we have $a_{i}=0$ and $m_1+m_2+1 \geq k$.

Since the length between $a_{i}$ and $b_{i}$ is $k$, we have two cases for $m_2$ and $m_3$:
\begin{itemize}
    \item If $m_{2}+m_{3}<k$, then $m_{2}<(k-1)$ and $m_{3}<(k-1)$, which means that the substring between $a_{i}$ and $b_{i}$ must contain at least one non-zero symbol.
    \item If $m_{2}+m_{3}\geq k$, then $m_{2}=m_{3}=(k-1)$, which means that the substring between $a_{i}$ and $b_{i}$ is $0^{k-1}$.
\end{itemize}

\emph{A) Descendants with $m_{2}+m_{3}<k$:} Since the substring between $a_{i}$ and $b_{i}$ must contain at least one non-zero symbol, the changes in $\mu(\bz')$, as well as $\mu(\bz)$, caused by $a_{i}$ and $b_{i}$, can be analyzed independently. If the non-zero symbol is $d\in\Sigma_{q}^{+}$, with $a_{i}$ and $b_{i}$ on the left and right side respectively, the changes in $\mu(\bz')$ can be separately studied on the two sides of $d$. In the following, we use $0^{j-1}a0^{k-j}$ or $0^{t-1}a0^{k-t}$ to denote a substring of length $k$ with $\wt(0^{j-1}a0^{k-j})=\wt(0^{t-1}a0^{k-t})=1$, where $j,t\in [k]$ and $a\in\Sigma_{q}^{+}$.
\begin{enumerate}
    \item The changes on the left side of $d$ is caused by changing $a_{i}$. Since $a_{i} = 0$, then $a=a_{i}+a \neq 0$.
           \begin{enumerate}
                 \item \label{itm:ambi_a_ik}If $\Bigl\lfloor\dfrac{m_{1}+m_{2}+1}{k}\Bigr\rfloor> \Bigl\lfloor\dfrac{m_{1}}{k}\Bigr\rfloor$, the length before $d$ increases by $k$ and the substring $0^{j-1}a0^{k-j}$ is inserted in $\mu(\bz')$, before the symbol $d$.

                 \item \label{itm:ambi_a_ambi} If $\Bigl\lfloor\dfrac{m_{1}+m_{2}+1}{k}\Bigr\rfloor= \Bigl\lfloor\dfrac{m_{1}}{k}\Bigr\rfloor$, the length before $d$ stays the same and $0$ is substituted by $a$ at $a_{i}$.
        \end{enumerate}

    \item The changes on the right side of $d$ is caused by changing $b_{i}$.
         \begin{enumerate}
              \item If $b_{i} \neq 0$,
                    \begin{enumerate}
                        \item if $b_{i}-a = 0$,
                           \begin{enumerate}
                               \item \label{itm:ambi_b_dk}if $\Bigl\lfloor\dfrac{m_{3}+m_{4}+1}{k}\Bigr\rfloor> \Bigl\lfloor\dfrac{m_{4}}{k}\Bigr\rfloor$, the length of $\mu(\bz')$ after $d$ decreases by $k$ and a substring $0^{t-1}a0^{k-t}$ is deleted from $\mu(\bz')$.

                               \item \label{itm:ambi_b_01}if $\Bigl\lfloor\dfrac{m_{3}+m_{4}+1}{k}\Bigr\rfloor= \Bigl\lfloor\dfrac{m_{4}}{k}\Bigr\rfloor$, the length after $d$ stays the same and $a$ is substituted by $0$ at $b_i$.
                           \end{enumerate}
                        \item \label{itm:ambi_b_02} if $b_{i}-a \neq 0$, the length after $d$ stays the same and $b_{i}$ is substituted by $(b_{i}-a)$.
                    \end{enumerate}
              \item If $b_{i} = 0$, then $b_{i}-a \neq 0$.
                    \begin{enumerate}
                         \item \label{itm:ambi_b_ik}if $\Bigl\lfloor\dfrac{m_{3}+m_{4}+1}{k}\Bigr\rfloor> \Bigl\lfloor\dfrac{m_{4}}{k}\Bigr\rfloor$, the length of $\mu(\bz')$ after $d$ increases by $k$ and the substring $0^{t-1}(0-a)0^{k-t}$ is inserted in $\mu(\bz')$.

                         \item \label{itm:ambi_b_03}if $\Bigl\lfloor\dfrac{m_{3}+m_{4}+1}{k}\Bigr\rfloor= \Bigl\lfloor\dfrac{m_{4}}{k}\Bigr\rfloor$, the length after $d$ stays the same and $0$ is substituted by $(0-a)$ at $b_{i}$.
                    \end{enumerate}
          \end{enumerate}
\end{enumerate}

Since $\bmu=\mu(\bz)$ and $\mu(\bz)=\mu(\bz')$, the changes from $\bmu=\mu(\bz')$ to $\bmu''=\mu(\bz'')$ are shown in Table~\ref{Table:specific_case} classified based on $a_i$ and $b_i$.

\begin{table*}
\begin{center}
   \caption{The changes in $\mu(\bz)$ with $m_2+m_3<k$.} \label{Table:specific_case}
        \begin{tabular}{p{3 cm}|p{2cm}|p{5cm}}
         \hline
           $a_{i}$ and $b_{i}$  & $|\bmu''|-|\bmu|$  &  $\bmu\to \bmu''$ \\ \hline
         \ref{itm:ambi_a_ik} and \ref{itm:ambi_b_dk}  & $0$ & insert $0^{j-1}a0^{k-j}$ and delete $0^{t-1}a0^{k-t}$ \\ \hline
         \ref{itm:ambi_a_ik} and \ref{itm:ambi_b_01}  & $+k$ & insert $0^{j-1}a0^{k-j}$ and $a\to0$ \\ \hline
        \ref{itm:ambi_a_ik} and \ref{itm:ambi_b_02}  & $+k$ & insert $0^{j-1}a0^{k-j}$ and $b_i \to (b_i-a)$\\ \hline
        \ref{itm:ambi_a_ik} and \ref{itm:ambi_b_ik}  & $+2k$ & insert $0^{j-1}a0^{k-j}$ and  $0^{t-1}(q-a)0^{k-t}$\\ \hline
        \ref{itm:ambi_a_ik} and \ref{itm:ambi_b_03}  & $+k$ & insert $0^{j-1}a0^{k-j}$ and  $0\to (0-a)$\\ \hline
        \ref{itm:ambi_a_ambi} and \ref{itm:ambi_b_dk}  & $-k$ & $0\to a$ and delete $0^{t-1}a0^{k-t}$ \\ \hline
         \ref{itm:ambi_a_ambi} and \ref{itm:ambi_b_01}  & $0$ & two substitutions ($0\to a$ and $a\to 0$) \\ \hline
        \ref{itm:ambi_a_ambi} and \ref{itm:ambi_b_02}  & $0$ &  two substitutions($0\to a$ and $b_i\to (b_i-a)$)\\ \hline
        \ref{itm:ambi_a_ambi} and \ref{itm:ambi_b_ik}  & $+k$ &$0\to a$ and  insert $0^{t-1}(0-a)0^{k-t}$\\ \hline
        \ref{itm:ambi_a_ambi}  and \ref{itm:ambi_b_03}  & $0$ & two substitutions($0\to a$ and $0\to (0-a)$)\\ \hline
        \end{tabular}

    \end{center}
\end{table*}

\emph{B) Descendants with $m_{2}+m_{3}>k$:} Based on the analysis above, when $m_{2}+m_{3}>k$, the substring between $a_{i}$ and $b_{i}$ is $0^{k-1}$. Hence $\bz'$ can be rewritten as
\[ \bz'= \bu 0^{m_{1}}a_{i}0^{k-1}b_{i}0^{m_{4}}\bv,\]
where $\bu,\bv \in \Sigma^{*}_{q}$. After one substitution, $\bz''$ can be expressed as
\[ \bz''= \bu \underline{0^{m_{1}}(a_{i}+a)0^{k-1}(b_{i}-a)0^{m_{4}}}\bv,\]
where $a_{i}=0$ and $a\in \Sigma_{q}^{+}$. Since the length of $\mu(\bz')$ is influenced by the \emph{underlined substring} above, we focus on the changes of this segment.

The length of the underlined substring satisfies
\[\Bigl\lfloor\dfrac{m_{1}+m_{4}+k+1}{k}\Bigr\rfloor= \Bigl\lfloor\dfrac{m_{4}}{k}\Bigr\rfloor+\Bigl\lfloor\dfrac{m_{1}}{k}\Bigr\rfloor+1,\]
or \[\Bigl\lfloor\dfrac{m_{1}+m_{4}+k+1}{k}\Bigr\rfloor= \Bigl\lfloor\dfrac{m_{4}}{k}\Bigr\rfloor+\Bigl\lfloor\dfrac{m_{1}}{k}\Bigr\rfloor+2.\]
The two cases are discussed below in detail.

If the length of the underlined substring satisfies  $\Bigl\lfloor\dfrac{m_{1}+m_{4}+k+1}{k}\Bigr\rfloor= \Bigl\lfloor\dfrac{m_{4}}{k}\Bigr\rfloor+\Bigl\lfloor\dfrac{m_{1}}{k}\Bigr\rfloor+1$, then the changes in $\mu(\bz')$ consist of two cases (based on the change from $(a_{i},b_{i})$ to $(a_{i}+a,b_{i}-a)$):
\begin{enumerate}
    \item if $(a_{i},b_{i})=(0,q_{i})$ with $q_i\neq 0$, then we again have two cases:
         \begin{enumerate}
             \item \label{enum:pk_10_no0_k} if $a_{i}+a,b_{i}-a$ are non-zero, the length of $\mu(\bz')$ increases by $k$, and the substring $0^{j-1}a0^{k-j}$ is inserted in $\mu(\bz')$ and $b_{i}$ is substituted by $b_{i}-a$.
             \item  \label{enum:primal_k}  if $(a_{i}+a,b_{i}-a)=(q_{i},0)$, we have $\mu(\bz'')=\mu(\bz')$.
         \end{enumerate}
    \item \label{enum:pk_20_no_k}  if $(a_{i},b_{i})=(0,0)$, then $a_{i}+a,b_{i}-a$ are non-zero, the length of $\mu(\bz')$ increases by $k$, and the substring $0^{j-1}a0^{k-j}$ is inserted in $\mu(\bz')$ and $0$ is substituted by $(0-a)$ at $b_i$.
\end{enumerate}

Similarly, if the length of the underlined substring satisfies  $\Bigl\lfloor\dfrac{m_{1}+m_{4}+k+1}{k}\Bigr\rfloor= \Bigl\lfloor\dfrac{m_{4}}{k}\Bigr\rfloor+\Bigl\lfloor\dfrac{m_{1}}{k}\Bigr\rfloor+2$, the changes in $\mu(\bz')$ also contain two cases:
\begin{enumerate}
    \item if $(a_{i},b_{i})=(0,q_{i})$, then there are two different cases:
         \begin{enumerate}
             \item \label{enum:pk_10_no0_2k} if $a_{i}+a,b_{i}-a$ are non-zero, the length of $\mu(\bz')$ increases by $k$, and the substring $0^{j-1}a0^{k-j}$ is inserted in $\mu(\bz')$ and $b_{i}$ is substituted by $b_{i}-a$.
             \item \label{enum:primal_2k}  if $(a_{i}+a,b_{i}-a)=(q_{i},0)$, we have $\mu(\bz'')=\mu(\bz')$.
         \end{enumerate}
    \item \label{enum:pk_20_no_2k} if $(a_{i},b_{i})=(0,0)$, then $a_{i}+a,b_{i}-a$ are non-zero, the length of $\mu(\bz')$ increases by $2k$, and the string $0^{j-1}a0^{k-j}$ and $0^{t-1}(0-a)0^{k-t}$ are inserted in $\mu(\bz')$
\end{enumerate}

Since the $k$-TDs do not change the duplication root, we have $\rt(\bx)=\rt(\bx')$ and $\mu(\bz)=\mu(\bz')$. Based on the analysis above, the changes in $\mu(\bz)$ caused by one substitution can be divided into four different cases:
\begin{itemize}
    \item if $|\mu(\bz'')|=|\mu(\bz)|+2k$, then $\mu(\bz'')$ is derived from $\mu(\bz)$ by inserting one $0^{j-1}a0^{k-j}$ and one $0^{t-1}(0-a)0^{k-t}$. Furthermore, $a$ and $(0-a)$ have distance $k$.

    \item if $|\mu(\bz'')|=|\mu(\bz)|+k$, then $\mu(\bz'')$ is derived from $\mu(\bz)$ by either inserting $0^{j-1}a0^{k-j}$ and substituting $b_i\to (b_i-a)$ or inserting $0^{t-1}(0-a)0^{k-t}$ and substituting $0\to a$. In both cases, $a$ and $(b_i-a)$ have a distance of $k$.

    \item if $|\mu(\bz'')|=|\mu(\bz)|$, three different cases occur. First, $\mu(\bz'')=\mu(\bz)$, there are no changes. Second, $\mu(\bz'')$ is derived from $\mu(\bz)$ by two substitutions ($0\rightarrow a$ and $b_i\rightarrow (b_i-a)$ with distance $k$). Third, the string $0^{j-1}a0^{k-j}$ is inserted and $0^{t-1}a0^{k-t}$ is deleted, where $a$ stays in the same position. In the third case, $\mu(\bz'')$ is derived from $\mu(\bz)$ by swapping $0^{e}$ with a substring (the form of $d$ or $c\Sigma_q^{e-2}d$ with $e\neq 0$ and $c,d\in \Sigma_{q}^{+}$) between $a_i=0$ and $b_i=a$, where the distance of the begining of the two substrings is $k$. Furthermore, the integer $e$ satisfies $1\leq e\leq (k-1)$.

   \item if $|\mu(\bz'')|=|\mu(\bz)|-k$, $\mu(\bz'')$ is derived from $\mu(\bz)$ by deleting $0^{t-1}a0^{k-t}$ and substituting $0\to a$.
 \end{itemize}

In conclusion, the changes from $\bmu=\mu(\bz)$ to $\bmu''=\mu(\bz'')$ caused by one substitution are described in the first and second columns of Table~\ref{Table:descendant_change_case}. We now discuss the changes in $\bmu_j$, i.e., the difference between $\bmu_j$ and $\bmu''_j$ for $j\in[k]$. This is done by considering four cases:

\deleted{
Based on the analysis above, after many exact $k$-TDs and one substitution, the string $\mu(\bz)$ is changed to $\mu(\bz'')$. Since we want to recover the duplication root $\rt(\bx)$, we need to recover the $\mu(\bz)$. However, it is a great challenge to recover $\mu(\bz)$ based on the $\mu(\bz'')$ directly.

According to Table~\ref{Table:descendant_change_case}, the length $|\mu(\bz)|$ of $\mu(\bz)$ changes at units of $k$, and the two substitutions ($0\to a$ and $b_i\to (b_i-a)$) have a distance of $k$. To make full use of the two properties, we split the $\mu(\bz)$ into $k$ strings. Based on Theorem~\ref{Theo:descendant_changes}, we obtain $\{\bmu_{1},\dotsc,\bmu_{k}\}$. Since the mappings between $\{\bmu_{1},\dotsc,\bmu_{k}\}$ and $\mu(\bz)$ are one-to-one, we can recover $\mu(\bz)$ if $\{\bmu_{1},\dotsc,\bmu_{k}\}$ are obtained.

In order to recover $\{\bmu_{1},\dotsc,\bmu_{k}\}$, we need to obtain all the changes of the $k$ strings. Based on Table~\ref{Table:descendant_change_case}, the changes of $\{\bmu_{1},\dotsc,\bmu_{k}\}$ can also be discussed in four cases:}
\begin{itemize}
    \item   If $|\mu(\bz'')|=|\mu(\bz)|+2k$, $\mu(\bz'')$ is derived from $\mu(\bz)$ by inserting a $0^{j-1}a0^{k-j}$ and a $0^{t-1}(0-a)0^{k-t}$. \deleted{For $\{\bmu_{1},\dotsc,\bmu_{k}\}$,} For $j\in[k]$, the length of each $\bmu_j$ increases by $2$. For one value of $j$, $a(0-a)$ is inserted in $\bmu_j$ and two $0$s are inserted in the other $(k-1)$ strings with a distance at most $2$.

    \item   If $|\mu(\bz'')|=|\mu(\bz)|+k$, $\mu(\bz'')$ is derived from $\mu(\bz)$ by inserting $0^{j-1}a0^{k-j}$ or $0^{t-1}(0-a)0^{k-t}$ and substituting $\left(b_i\rightarrow (b_i-a)\right)$ or $(0 \rightarrow a)$. For $j\in [k]$, the length of $\bmu_j$ increases by $1$. For one value of $j$, the insertion and substitution $b_{i}\rightarrow a(b_{i}-a)$ occur in $\bmu_j$ and $0$ is inserted into each of the other $(k-1)$ strings.

    \item   If $|\mu(\bz'')|=|\mu(\bz)|$, $\mu(\bz'')$ is derived from $\mu(\bz)$ in three different cases. First, $\mu(\bz'')=\mu(\bz)$, there are no changes. Second,  $\mu(\bz'')$ is derived from $\mu(\bz)$ by substituting two symbols $(0 \rightarrow a, b_{i} \rightarrow (b_{i}-a))$ with distance $k$. For one value of $j$, the substitutions $(0 b_{i} \rightarrow a(b_i-a))$ occur in $\bmu_j$ and the other $(k-1)$ strings stay the same. Third, $\mu(\bz'')$ is obtained from $\mu(\bz)$ by inserting $0^{j-1}a0^{k-j}$ and deleting $0^{t-1}(0-a)0^{k-t}$. For $j\in[k]$, at least one $\bmu_j$ swaps  $(b0) \to (0b)$ with $b\in \Sigma_{q}^{+}$ and the other strings stay the same.

    \item  If $|\mu(\bz'')|=|\mu(\bz'')|-k$, $\mu(\bz'')$ is derived from $\mu(\bz)$ by deleting $0^{t-1}a0^{k-t}$ and substituting $0\rightarrow a$. For $\{\bmu_{1},\dotsc,\bmu_{k}\}$, one $0$ is deleted from each of the $k$ strings.
\end{itemize}

The changes of $\{\bmu_{1},\dotsc,\bmu_{k}\}$ can be summarized in the third column of Table~\ref{Table:descendant_change_case}. The forth column is obtained by noting that $\bs_j=\Gamma(\bmu_j),j\in[k]$. This completes the proof of Table~\ref{Table:descendant_change_case}.

\deleted{
Based on the weight-indicating rule in~\eqref{eq:weig_indicating_map} and the changes of $\{\bmu_{1},\dotsc,\bmu_{k}\}$, the changes of $\{\bs_{1},\dotsc,\bs_{k}\}$ are summarized in the forth column of  Table~\ref{Table:descendant_change_case}. Combing the analysis above together, we prove the Table~\ref{Table:descendant_change_case} completely.}

Second, we consider the case in which the substitution position $p$ satisfies $(|\bx'|-k)<p\leq |\bx'|$, which means that one symbol in $\bz$ changes. Since one substitution only changes one symbol in $\bz'$, we have
  \[ \bz''= \bu a_{1}a_{2} \cdots (a_{i}+a)\cdots a_{k}.\]
where $a\in \Sigma_{q}^{+}$. Since the substitution only occurs in a tandem duplication copy, we have $a_{i}=0$ and $m_1+m_2+1 \geq k$. Note that $a=a_{i}+a \neq 0$. There are two cases to consider:
\begin{enumerate}
                 \item If $\Bigl\lfloor\dfrac{m_{1}+m_{2}+1}{k}\Bigr\rfloor> \Bigl\lfloor\dfrac{m_{1}}{k}\Bigr\rfloor$, then the length of $\mu(\bz')$ increases by $k$ and the substring $0^{j-1}a0^{k-j}$ is inserted into $\mu(\bz')$.
                 \item If $\Bigl\lfloor\dfrac{m_{1}+m_{2}+1}{k}\Bigr\rfloor= \Bigl\lfloor\dfrac{m_{1}}{k}\Bigr\rfloor$, then the length of $\mu(\bz')$ stays the same and $0$ is substituted by $a$ at $a_i$.
\end{enumerate}

We can then find the difference between $\bmu_j$ and $\bmu''_j$, and $\bs_j$ and $\bs''_j$, $j\in[k]$, which are listed in Table~\ref{Table:descendant_change_case_tail}. This completes the proof of Theorem~\ref{Theo:descendant_changes}.

\deleted{
Since the $k$-TDs errors do not change the duplication root, we have $\rt(\bx)=\rt(\bx')$ and $\mu(\bz)=\mu(\bz')$. Based on the analysis above, the changes in $\mu(\bz)$ with $(|\bx'|-k)<p\leq |\bx''|$ are summarized in the first two columns of Table~\ref{Table:descendant_change_case_tail}. Similarly, we obtain the changes in $\{\bmu_{1},\dotsc,\bmu_{k}\}$ and $\{\bs_{1},\dotsc,\bs_{k}\}$ in the third and forth columns of Table~\ref{Table:descendant_change_case_tail}.

In conclusion, after analyzing the changes in $\mu(\bz)$ based on different substitution position $p$, we prove Theorem~\ref{Theo:descendant_changes}.}
\end{IEEEproof}

\section{The proof of Theorem~\ref{Theo:Binary_code}}
\begin{IEEEproof}
 Given a codeword $\bs\in C_{(a,b,c)}$, let $\bs''$ be obtained from $\bs$, either with no error, or via one of the errors listed in Theorem~\ref{Theo:Binary_code}.
\begin{enumerate}
    \item If $|\bs''|=|\bs|-1$, then there has been a single deletion, correctable via the VT code~\eqref{eq:binary_VT}.

    \item If $|\bs''|=|\bs|$, then there are the following possibilities: no error, a single substitution, swapping two adjacent different symbols, $00\to11$, and $11\to00$.  
    Based on \eqref{eq:binary_weight_change}, we have $\sum^{n}_{i=1} s''_i=(b+b'') \bmod 5$, and $b''$, along with the syndrome of the VT code, is helpful for distinguishing these cases. If $b''=2$, one substitution $00\to11$ between $\bs$ and $\bs''$ has occurred. We have $\sum_i i s''_i=a+2p+1 \bmod (2n+3)$, where $p$ is the position of the substitution. Hence, we can recover $\bs$ by one substitution $11\to 00$ at the position $p$ of $\bs''$. If $b''=-2$, a substitution $11\to 00$ has occurred from $\bs$ to $\bs''$. We have $\sum_i i s''_i=a-2p-1 \bmod (2n+3)$. Then we can recover $\bs$ from $\bs''$ by flipping two symbols at positions $p$ and $p+1$. If $b''=1$, a substitution $0\to1$ has occurred. We have $\sum_i i s''_i=a+p \bmod (2n+3)$. Hence, we can recover $\bs$ by one substitution $1\to 0$ at position $p$ of $\bs''$. If $b''=-1$, a substitution $1\to 0$ has occurred. We have $\sum_i i s''_i=a-p \bmod (2n+3)$. Then $\bs$ can be recovered by flipping the symbol in the $p$th position of $s''$. If $b''=0$ and the VT syndrom has changed, an adjacent transposition has occurred in $\bs$. If the transposition occurs at $p$, for the constructed string $\{\bs^{cs}|s^{cs}_{i}=\sum^{i}_{j=1} s_j,i\in [|\bs|]\}$, the string $\bs^{cs}$ and $\bs^{cs''}$ only differ at position $p$ with $|s^{cs}_{p}-s^{cs''}_{p}|=1$~\cite{gabrys2017codes}. Then we have $\sum_i i \left( \sum^{i}_{j=1} s''_j\right)=c\pm p \bmod (2n+1)$. Thus, we can recover $\bs$ by swapping the two symbols at positions $p$ and $(p+1)$ of $\bs''$. 

    \item If $|\bs''|=|\bs|+1$, based on Theorem~\ref{Theo:descendant_changes}, $\bs''$ is derived from $\bs$ in one of the following ways: inserting a $0$, inserting a $1$, $0\to 11$, or $1\to00$. Based on \eqref{eq:binary_VT} and \eqref{eq:binary_weight_change}, we have $\sum_i i s''_i=(a+a'') \bmod (2n+3)$ and $\sum_i  s''_i=(b+b'') \bmod 5$.   If $b''=0$ and $a''\leq \wt(\bs'')$, one $0$ is inserted in $\bs$, and we can recover $\bs$ by deleting it \cite{sloane2000single}. 
    If $a''> \wt(\bs'')$ and $b''=1$, one $1$ is inserted in $\bs$. Then we can recover $\bs$ by deleting a $1$ from $\bs''$~\cite{sloane2000single}. If $a''> \wt(\bs'')$ and $b''=2$, $\bs''$ is derived from $\bs$ by 
    a substitution $0\to 11$. We have $a''=2p+1+r_{1}$, where $p$ denotes the position of the original 0 and $r_1$ denotes the number of $1$s on its right. 
    During the recovery process, we denote our guess for the position and the number of $1$s on the right side of the position as $p'$ and $r'_1$, respectively. If $r'_1<r_1$, then $2p'+1+r'_1>2p+1+r_1$. If $r'_1>r_1$, then $2p'+1+r'_1<2p+1+r_1$. Only if $r'_1=r_1$, then $2p'+1+r'_1=2p+1+r_1$. 
    If $b''=-1$, the substitution $1\to 00$ has occurred. If the substitution is at the position $p$, then $a''=-p+r_1$, where $r_1$ denotes the number of $1$s on the right side of the substitution. Similar to correcting the substitution $0\to 11$, we can obtain the position $p$ and recover $\bs$ by the substitution $00\to 1$ at the positions $p,(p+1)$ of $\bs''$.

    \item If $|\bs''|=|\bs|+2$, then $\bs''$ is derived from $\bs$ in one of three ways: inserting $11$, inserting $00$, or inserting two $0$s separated by $1$ ($1\to 010$). Based on \eqref{eq:binary_VT} and \eqref{eq:binary_weight_change}, we have $\sum_i i s''_i=(a+a'') \bmod (2n+3)$ and $\sum_i  s''_i=(b+b'') \bmod 5$. If $b''=2$, $11$ is inserted in $\bs$.  Let $p$ denote the position in which $11$ is inserted. Based on \eqref{eq:binary_VT}, we have $a''=(p+p+1)+2r_1=2(l_0+l_1+1)+1+2r_1=2(l_1+r_1+2)+2l_0-1=2\wt(\bs'')+2l_{0}-1$, where $l_1$ and $r_1$ denote the number of $1$s at the left and right sides of the position $p$, and $l_0$ denotes the number of $0$s at the left side of the inserting position. Then we can recover $\bs$ by deleting one $11$ from $\bs''$ after $l_0$ $0$s from the beginning. If $b''=0$, two $0$s are inserted in $\bs$. If $a''=0 \bmod 2$, $00$ is inserted in $\bs$ and $a''=2r_1$, where $r_1$ denotes the number of $1$s on the right side of the insertion position. Then we can recover $\bs$ by deleting $00$ from $\bs''$ before $r_1$ $1$s from the end of $\bs''$. If  $a''=1 \bmod 2$, two $0$s are inserted in $\bs$ separated by $1$ and $a''=2r_1+1$. Similarly, we can recover $\bs$ by deleting two $0$s before $r_1$ and $r_1+1$ $1$s from the end of $\bs''$.
\end{enumerate}

These error patterns include all those occurring in
 $\{\bs_{1},\dotsc, \bs_{k}\}$ caused by many exact $k$-TDs and at most one substitution error in the noisy duplication channel.
\end{IEEEproof}

\section{The proof of Theorem~\ref{Theom:nd_error_code}}
\begin{IEEEproof}
To prove Theorem~\ref{Theom:nd_error_code}, we have to show that the error-correcting code $C_{nd}$ in Construction~\ref{con:noisy_duplication_code} can correct all error patterns in $\{\bmu_{1},\dotsc,\bmu_{k}\}$. Based on Theorem~\ref{Theo:Binary_code}, the code $C_{(a,b,c)}$ over $\Sigma_{2}$ can correct all error patterns shown in the $\bmu_{j}$ column of Tables~\ref{Table:descendant_change_case} and \ref{Table:descendant_change_case_tail} in rows marked by $(\$)$. The constraints~\eqref{eq:nd_Tq_odd}, \eqref{eq:nd_Tq_even},  \eqref{eq:nd_Tq_Culsum} and \eqref{eq:nd_Tq_concatinating} can correct the other error patterns.

Given a codeword $\bx\in C_{nd}\subseteq \ir(n)\cap \Sigma_{q}^{n}$, we have $\phi(\rt(\bx))=(\by,\bmu)$ with $\by=\hat\phi(\bx)\in\Sigma_{q}^{k}$ and $\bmu=\mu(\bz)=\bz=\bar \phi(\bx)\in \Sigma_{q}^{n-k}$. After many exact $k$-TDs and at most one substitution, we obtain a descendant $\bx''\in D_{k}^{*(\leq 1)}(\bx)$ with $\phi(\bx'')=(\by, \bz'')$ and $\bz''=\bar\phi(\bx'')$. In the following, we can recover the codeword $(\by,\bmu)$ by correcting four types of error patterns in $(\by,\bmu'')$, where $\bmu''=\mu(\bz'')$. Based on the recovered $(\by,\bmu)$, we can obtain the duplication root $\rt(\bx)$ and thus the codeword $\bx$. The four cases are below:
\begin{itemize}
    \item If $|\bmu''|=|\bmu|-k$, then a $0$ is deleted from both $\{\bmu_{1},\dotsc,\bmu_{k}\}$ and $\{\bs_{1},\dotsc,\bs_{k}\}$. By \eqref{eq:nd_VT}, we recover $\{\bs_{1},\dotsc,\bs_{k}\}$ by inserting a $0$ in each of them. Based on \eqref{eq:weig_indicating_map}, the positions of $0$s between $\{\bmu_{1},\dotsc,\bmu_{k}\}$ and $\{\bs_{1},\dotsc,\bs_{k}\}$ coincide. We can recover $\{\bmu_{1},\dotsc,\bmu_{k}\}$ by inserting $0$s at the same positions in $\{\bs_{1},\dotsc,\bs_{k}\}$.

    \item If $|\bmu''|=|\bmu|$, $\{\bmu_{j},j\in[k]\}$ contain two types of errors: transpositions of $0$ and $b$ in more than one $\bmu_j$, or the substitution either $0c\to a(c-a)$ or $0 \to a$ in one $\bmu_j$. By \eqref{eq:nd_VT}, we have
    \[\sum^{|\bs''_{j}|}_{i=1}is''_{ji}=(a_{j}+a''_{j})\bmod (2|\bs_{j}|+3),\quad j\in[k].\]
    If $\{a''_{j}, j\in[k]\}$ contain more than one non-zero integer, both $\{\bmu_{j},j\in[k]\}$ and $\{\bs_{j},j\in [k]\}$ with non-zero $\{a''_{j},j\in[k]\}$ contain one adjacent transposition of $(0,b)$ and $(0,1)$, respectively. By \eqref{eq:binary_transition}, the transposition positions $\{p_{j},j\in [k]\}$ can be obtained. Since both $\{\bmu_{j},j\in[k]\}$ and $\{\bs_{j},j\in [k]\}$ contain adjacent transpositions at the same positions, we can recover $\{\bmu_{j},j\in[k]\}$ by swapping two symbols starting at $\{p_{j},j\in [k]\}$. If $\{a''_{j}, j\in[k]\}$ only contain one non-zero integer, say $a''_{1}$, three types of errors may occur based on the weight change of $\{\bs_{j},j\in [k]\}$ by \eqref{eq:nd_weight}. Based on the proof of Theorem~\ref{Theo:Binary_code}, we can obtain the change position $p_1$ in $\bmu_{1}$ and $\bs_{1}$. If $p_1<|\bmu_{1}|$, according to Table~\ref{Table:descendant_change_case}, $\bmu_1$ contains one substitution $0c \to a(c-a)$, we can recover $\bmu_{1}$ by the substitution $\mu'_{1p_1}\mu'_{1(p_1+1)}\to 0(\mu'_{1p_1}+\mu'_{1(p_1+1)})$. If $p_1=|\bmu_{1}|$, according to Table~\ref{Table:descendant_change_case_tail}, $\bmu_1$ contains one substitution $0 \to a$, we can recover $\bmu_{1}$ by the substitution $\mu'_{1p_1}\to 0$.

    \item If $|\bmu''|=|\bmu|+k$, then $(k-1)$ of $\{\bmu_{j},j\in[k]\}$ contain one insertion $\Lambda \to 0$, and one string, say $\bmu_{k}$, contains either one insertion $\Lambda\to a$ in Table~\ref{Table:descendant_change_case_tail} or one insertion and one substitution $c\to a(c-a)$ in Table~\ref{Table:descendant_change_case}. By \eqref{eq:binary_VT}, the $(k-1)$ strings $\{\bmu_{j},j\in[k-1]\}$ can be recovered. After that, we generate $\inl'(\bmu)=\bmu_{1}\cdots\bmu_{(k-1)}\bmu'_{k}$ by concatenating the $k$ strings. Compared to $\inl(\bmu)$, $\inl'(\bmu)$ contains either one insertion $\Lambda\to a$ or one insertion and one substitution $c\to a(c-a)$. Based on \eqref{eq:nd_Tq_odd}, \eqref{eq:nd_Tq_even} and Construction~\ref{con:Tq_Code}, we obtain the changes $(\Delta \bar a_1,\Delta \bar a_2)$. If $\Delta \bar a_1+\Delta \bar a_2 \neq 0 \bmod q$, then $\inl'(\bmu)$ contains one insertion $\Lambda\to a$. Then we can recover the insertion $\Lambda\to a$ by \eqref{eq:nd_Tq_concatinating}. If $ \Delta \bar a_1+\Delta \bar a_2=0 \bmod q$,   $\inl'(\bmu)$ contains one insertion and one substitution $c\to a(c-a)$.
    \deleted{
    if only one of $(\bar a_1,\bar a_2)$ changes, then $\inl'(\bmu)$ contains one insertion $\Lambda\to a$. Since both \eqref{eq:nd_Tq_odd}
    and \eqref{eq:nd_Tq_even} can correct one insertion in $\Od(\inl(\bmu))$ and $\Even(\inl(\bmu))$, we can recover $\inl(\bmu)$ and $\{\bmu_{j},j\in[k]\}$.
    If $(\bar a_1,\bar a_2)$ both change, }
    By \eqref{eq:cumulative_sum} and the fact that $a+(c-a)=c$, we construct $r'=\cusum(\inl'(\bmu))$ with one insertion. Since \eqref{eq:nd_Tq_Culsum} can correct one insertion in $\cusum(\inl'(\bmu))$, we can recover $\cusum(\inl(\bmu))$, $\inl(\bmu)$, and $\{\bmu_{j}, j\in[k]\}$.

    \item If $|\bmu''|=|\bmu|+2k$, then $(k-1)$ strings of $\{\bmu_{j}, j\in[k]\}$ insert two $0$s with distances at most $2$, and one string such as $\bmu_{1}$ contains one insertion $a(0-a)$. Similar to the proof of Theorem~\ref{Theo:Binary_code}, based on \eqref{eq:nd_VT}, we can recover $\{\bmu_{2},\dotsc,\bmu_{k}\}$ by deleting two $0$s. After that, we generate the string $\inl'(\bmu)=\bmu'_{1}\bmu_{2}\cdots \bmu_{k}$. Obviously, the string $\inl'(\bmu)$ contains one insertion $a(0-a)$. When $\inl(\mu(\bz))$ is divided into two strings $\Od(\inl(\mu(\bz)))$ and $\Even(\inl(\mu(\bz)))$, one symbol is inserted into each of $\Od(\inl(\mu(\bz)))$ and $\Even(\inl(\mu(\bz)))$ to generate $\Od(\inl'(\mu(\bz)))$ and $\Even(\inl'(\mu(\bz)))$. Since both~\eqref{eq:nd_Tq_odd}
    and \eqref{eq:nd_Tq_even} can correct an insertion of one symbol in $\Od(\inl(\bmu))$ and $\Even(\inl(\bmu))$, respectively, we can recover $\inl(\bmu)$ and $\{\bmu_{j}, j\in[k]\}$.
\end{itemize}

Having recovered $\{\bmu_{j}, j\in[k]\}$, we can reconstruct $\bmu$, the duplication root $\rt(\bx)$, and the codeword $\bx\in C_{nd}$. Thus, the error-correcting code $C_{nd}$ can correct all the error patterns caused by many exact $k$-TD and at most one substitution.

\deleted{
Because the integers $j, a_j, c_j,b,\bar a_1,\bar a_2, \bar a_3,\bar b_1,\bar b_2, \bar b_3$ can be any value in their corresponding ranges, the number of possible codes is $5q^{3}\lceil\frac{n-k}{2}\rceil^2(2\lceil\frac{n-k}{k}\rceil+3)^k(2\lceil\frac{n-k}{k}\rceil+1)^{k}(n-k)$. These codes partition the set $\ir(n)$, so there is at least one code with size
   \deleted{Furthermore, since the code $C_{nd}$ satifies $C_{nd}\in \ir(n)\cap\Sigma_{q}^{n}$, the size of candidate strings for a single code is $|\ir(n)|$. As a result, we can find at least one code $C_{nd}$ with code size }
\[|C_{nd}|\geq\frac{|\ir(n)|}{5q^{3}\lceil\frac{n-k}{2}\rceil^2(2\lceil\frac{n-k}{k}\rceil+3)^k(2\lceil\frac{n-k}{k}\rceil+1)^{k}(n-k)}.\]

Since $\lceil\frac{n-k}{k}\rceil=\lceil\frac{n}{k}\rceil-1$, the code size of $C_{nd}$ can be rewritten as
\[|\ir(n)|\geq|C_{nd}|\geq\frac{|\ir(n)|}{5q^{3}\lceil\frac{n-k}{2}\rceil^2(4\lceil\frac{n}{k}\rceil^2-1)^{k}(n-k)}.\]}

Because the integers $j, a_j, c_j,b,\bar a_1,\bar a_2, \bar a_3, \bar a_4,\bar b_1,\bar b_2, \bar b_3, \bar b_4$ can be any value in their corresponding ranges, the number of possible codes is $5q^{3}\lceil\frac{n-k}{2}\rceil^2(2\lceil\frac{n-k}{k}\rceil+3)^k(2\lceil\frac{n-k}{k}\rceil+1)^{k}(n-k)$. These codes partition the set $\ir(n)$, so there is at least one code with size
   \deleted{Furthermore, since the code $C_{nd}$ satisfies $C_{nd}\in \ir(n)\cap\Sigma_{q}^{n}$, the size of candidate strings for a single code is $|\ir(n)|$. As a result, we can find at least one code $C_{nd}$ with code size }
\[|C_{nd}|\geq\frac{|\ir(n)|}{5q^{4}\lceil\frac{n-k}{2}\rceil^2(2\lceil\frac{n-k}{k}\rceil+3)^k(2\lceil\frac{n-k}{k}\rceil+1)^{k}(n-k)^{2}}.\]

Since $\lceil\frac{n-k}{k}\rceil=\lceil\frac{n}{k}\rceil-1$, the code size of $C_{nd}$ can be rewritten as
\[|\ir(n)|\geq|C_{nd}|\geq\frac{|\ir(n)|}{5q^{4}\lceil\frac{n-k}{2}\rceil^2(4\lceil\frac{n}{k}\rceil^2-1)^{k}(n-k)^{2}}.\]

%
%
\end{IEEEproof}

\end{appendices}

\end{document}